# AN OPTICAL and X–RAY STUDY of ABELL 576, a GALAXY CLUSTER WITH a COLD CORE[*]

Joseph J. Mohr[1,2,3], Margaret J. Geller[3], Daniel G. Fabricant[3], Gary Wegner[4], John Thorstensen[4] & Douglas O. Richstone[2]

## ABSTRACT

We analyze the galaxy population and dynamics of the galaxy cluster Abell 576; the observational constraints include 281 redshifts (230 new), $R$ band CCD galaxy photometry over a $2h^{-1}$ Mpc $\times$ $2h^{-1}$ Mpc region centered on the cluster, an *Einstein* IPC X–ray image, and an *Einstein* MPC X–ray spectrum. We focus on an 86% complete magnitude limited sample ($R_{23.5} < 17$) of 169 cluster galaxies.

The cluster galaxies with emission lines in their spectra have a larger velocity dispersion and are significantly less clustered on this $2h^{-1}$ Mpc scale than galaxies without emission lines. We show that excluding the emission line galaxies from the cluster sample decreases the velocity dispersion by 18% and the virial mass estimate by a factor of two.

The central cluster region contains a non–emission galaxy population and an intracluster medium which is significantly cooler ($\sigma_{core} = 387^{+250}_{-105}$ km/s and $T_X = 1.6^{+0.4}_{-0.3}$ keV at 90% confidence) than the global populations ($\sigma = 977^{+124}_{-96}$ km/s for the non–emission population and $T_X > 4$ keV at 90% confidence). Because (1) the low dispersion galaxy population is no more luminous than the global population and (2) the evidence for a cooling flow is weak, we suggest that the core of A576 may contain the remnants of a lower mass subcluster.

We examine the cluster mass, baryon fraction and luminosity function. The cluster virial mass varies significantly depending on the galaxy sample used. Consistency between the hydrostatic and virial estimators can be achieved if (1) the gas temperature at $r \sim 1 h^{-1}$ Mpc is $T_X \sim 8 keV$ (the best fit value) and (2) several velocity outliers are excluded from the virial calculation. Although the best fit Schechter function parameters and the ratio of galaxy to gas mass in A576 are typical of other clusters, the baryon fraction is relatively low. Using a lower limit to the cluster binding mass, we show that the gas mass fraction is $\leq 3h^{-3/2}$% and the baryon fraction is $\leq 6$%.

*Subject Headings:* galaxies: clusters: general — galaxies: evolution — galaxies: intergalactic

[1] Department of Physics, University of Michigan, Ann Arbor, MI 48109
[2] Department of Astronomy, University of Michigan, Ann Arbor, MI 48109
[3] Harvard–Smithsonian Center for Astrophysics, Cambridge, MA 02138
[4] Department of Physics and Astronomy, Dartmouth College, Hanover, NH 03755

medium — galaxies: luminosity function — galaxies: photometry — Xrays: galaxies

## 1. INTRODUCTION

The complex nature of galaxy clusters often makes it inappropriate to apply the simple equilibrium analyses which yield direct constraints on the character of dark matter. Statistical studies demonstrate that clusters are still forming at the present epoch (Geller & Beers 1982; Dressler & Shectman 1988; Jones & Forman 1992; Mohr *et al.* 1995). Detailed studies of particular clusters generally reinforce these results (e.g. Burns *et al.* 1994, 1995; Fabricant *et al.* 1986, 1989, 1993; Oegerle & Hill 1994; Sodré *et al.* 1992; Zabludoff & Zaritsky 1995). Although this commonplace complexity makes the straighforward application of dynamical models suspect, it does provide a cosmological clue. Analysis of the structure of large samples of clusters constrains the mean density of the universe (Richstone, Loeb & Turner 1992; Lacey & Cole 1993, Mohr *et al.* 1995).

This paper contains the results of the detailed study of the cluster Abell 576. We are drawn to study this cluster because of an apparent lack of substructure; the X–ray emission from A576 is symmetric and the original galaxy velocity sample (Melnick & Sargent 1977; Hill *et al.* 1980; Hintzen *et al.* 1982) provides little evidence for a recent merger. In particular, the centroid variation of the X–ray emission in Abell 576 is small ($w_{\vec{x}} = 12 \pm 4h^{-1}$ kpc) compared to the observed range in a sample of 65 clusters imaged with the *Einstein* Imaging Proportional Counter (IPC) (Mohr *et al.* 1995). The lack of substructure implies that application of equilibrium models in A576 may be more appropriate, allowing us to, for example, more accurately compare the cluster mass distribution to the distribution of gas and galaxies. With these goals in mind, we set out to augment the observational constraints on A576. We use multi–fiber spectroscopy to gather velocities for a large sample of cluster members and a mosaic of $R$ band CCD images to obtain accurate galaxy photometry over a large field. We combine these constraints on the cluster galaxies with the X–ray photometric and spectroscopic constraints on the cluster gas and dark matter distributions.

Although there is no compelling, new evidence of recent subcluster mergers in this dataset, the data do reveal a cluster which is far from the simple system appropriate for straightforward equilibrium analyses. In §2 we describe the acquisition and reduction of the optical data. Section 3 contains a discussion of two distinct galaxy populations: galaxies with emission lines in their spectra and galaxies without. We demonstrate the existence of a cool core in both the galaxy population and the intracluster medium (White & Silk 1980; Rothenflug *et al.* 1984) and then discuss two heuristic models in §4. Section 5 contains a calculation of the cluster luminosity function and the radial distribution of galaxies, a comparison of the hydrostatic and virial mass estimates, and a discussion of the $R$ band mass–to–light ratio and baryon fraction. We summarize our results in §6. Throughout the paper we take $H_0 = 100h$ km/s/Mpc.

## 2. OPTICAL DATA

Abell 576 is a richness class 1 Abell cluster (Abell 1958) centered on $\alpha$: 7:17.3 $\delta$: +55:50

(1950) at a radial velocity $cz = 11,600$ km/s. We have gathered an extensive optical data set consisting of an $R$ band CCD mosaic covering a $2h^{-1}$ Mpc $\times$ $2h^{-1}$ Mpc region (where $H_0 = 100h$ km/s/Mpc), and a sample of 281 redshifts within a projected distance of $\sim 1.5h^{-1}$ Mpc of the cluster center. Here we describe the observations, define the cluster velocity range and probe for substructure.

### 2.1. Observations and Data Reduction

Table 1 contains velocities (corrected to the solar system barycenter) and photometry of 281 galaxies within a $1.5° \times 1.5°$ ($3h^{-1}$ Mpc $\times$ $3h^{-1}$ Mpc) region centered on Abell 576. The columns contain the galaxy position, isophotal magnitude and uncertainty, line–of–sight velocity and uncertainty, and designation as emission (E) or non–emission (N). Fifty-one galaxy velocities come from ZCAT (Huchra et al. 1992); most of these are from Medusa spectroscopy (Hintzen et al. 1982) with quoted uncertainties of 100 km/s. Some of the Medusa velocities are reobservations of galaxies originally studied by Melnick and Sargent (1977).

We measured 230 velocities during the winters of '93 and '94 using the Decaspec (Fabricant & Hertz 1990) and the MkIII spectrograph mounted on the Michigan–Dartmouth–MIT 2.4m telescope. The Decaspec spectra have 12Å resolution with coverage from 4,500Å to 8,500Å. We reduce the spectra using the IRAF NOAO.TWODSPEC and RVSAO packages. Line–of–sight velocities of the galaxies with absorption line spectra are extracted by cross correlating with a template consisting of a combination of appropriately (zero–) shifted spectra from several stellar velocity standards. We use a line profile fitting procedure for those spectra with emission lines. The range of signal–to–noise in the spectra yields a median velocity uncertainty of 45 km/s. This uncertainty consists of two terms added in quadrature; the first term is the statistical uncertainty extracted from the cross correlation with template spectra (or from line profile fitting) and the second term is the dispersion solution uncertainty determined from the variation in the positions of 4 sky lines in the entire sample of sky spectra. For galaxies with velocities measured by line profile fitting, we add an additional 60 km/s uncertainty in quadrature because the emission line regions do not necessarily trace the galaxy center of mass (Thorstensen 1993, Kurtz et al. 1995).

We test the internal accuracy of our velocity measurements and the scale of our velocity uncertainties with multiple observations of 14 galaxies. The distribution of the absolute errors is acceptably small; the average of the absolute value of the velocity differences is 71 km/s or $0.94\sigma_v$ (where $\sigma_v$ is the velocity uncertainty for each observation). The largest velocity difference is 181 km/s, corresponding to a $2.5\sigma_v$ error.

During non–photometric conditions in Spring '94 and '95 we obtained an $R$ band CCD mosaic of the central $1° \times 1°$ ($2h^{-1}$ Mpc $\times$ $2h^{-1}$ Mpc) region of A576 with the Mt. Hopkins 1.2m. These thirty–six 5 min images are placed on a common zero–point using nine 1.5 min CCD exposures (each one overlaps with 4 of the deep images) taken in photometric conditions at the MDM 1.3m. We use Landolt (1992) standards to reduce these photometric images to the Johnson–Kron–Cousins system. The photometric solution has an RMS scatter of 0.012 mag. Examination of the curves of growth of several isolated standard stars reveals that our aperture photometry (8 arcsec radius) gathers between 96% and 98% of the total light in the stellar profile. A differential effect between galaxies and stars could lead to a systematic zeropoint error. We crudely examine this affect by comparing the effective isophotal radii ($r_{eff} = \sqrt{A/\pi}$ where $A$ is the isophotal area) for the galaxies to the stellar aperture. For galaxies brighter than $R_{23.5} = 16.5$, $r_{eff} \geq 8$ arcsec; for galaxies with $R_{23.5} = 19$, $r_{eff} \sim 4$ arcsec. Thus, for galaxies with $R_{23.5} = 16.5$, we expect minimal zeropoint errors, and for the brightest and dimmest galaxies in our sample we expect systematic zeropoint errors (of opposite sign) at the $\sim 3\%$ level. We measure the offset between the Johnson–Kron–Cousins system and each deeper non–photometric image with stellar aperture photometry of the $\sim 12$ isolated and unsaturated stars which appear in the deep image and the associated photometric image (we use the IRAF NOAO.APPHOT package). Finally, using the distribution of sky brightnesses and zero–points of the 36 deep images, we choose the $R = 23.5$ mag/arcsec$^2$ isophotal level which is $> 2.9\sigma$ above the noise per pixel in all images. The star galaxy separation and final isophotal $R_{23.5}$ galaxy photometry is performed using FOCAS (Jarvis & Tyson 1981; Valdes 1982). We review the star–galaxy separation for all non–stellar objects brighter than $R_{23.5} = 19$ to remove double stars which are typically misclassified as galaxies.

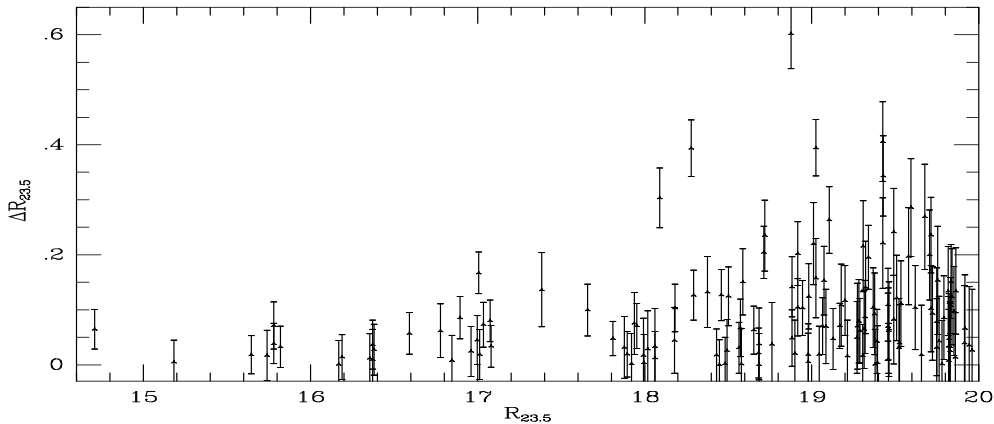

**Figure 1:** The absolute value of the measured magnitude differences and uncertainties in $R_{23.5}$ for 143 galaxies which appear on more than one deep image.

The magnitude uncertainties include three components added in quadrature; the first component is a 0.012 mag RMS scatter in the photometric solution, the second component is the Poisson noise and the third component is the uncertainty in the zero–point calculated using the photometric overlap images. The zero point uncertainty is the RMS scatter in the offsets calculated using the population of isolated stars the images have in common. The mean offset and the RMS scatter are all variance weighted values; the stars with the most accurate magnitudes carry more weight in determining the mean zero–point offset. For the galaxy sample with measured velocities, this approach yields magnitudes with minimum errors of $\sim 0.016$ mag and median errors of about 0.035 mag.

Figure 1 is a plot of the absolute value of the magnitude error as a function of $R_{23.5}$ magnitude for 143 galaxies with multiple magnitude measurements (the sign of the magnitude error contains no information in this setting). FOCAS produces reasonably accurate isophotal magnitudes except for galaxies which lie within regions of rapidly varying background (the haloes of other brighter galaxies or saturated stars). In these cases errors in the background determination produce significant magnitude errors ($\sim 0.25$ mag). This problem affects only a small portion of our sample ($< 5\%$).

For 51 galaxies with measured velocities located outside the CCD mosiac, we obtain $R$ magnitudes from a $1.5° \times 1.5°$ ($3h^{-1}$ Mpc $\times 3h^{-1}$ Mpc) digitized scan of the POSS E plate. To derive magnitudes, we use a sample of galaxies with measured $R_{23.5}$ magnitudes to determine the zero point in the scan magnitudes. The scatter in the mapping of scan magnitudes to $R_{23.5}$ is $\sim 0.25$ mag.

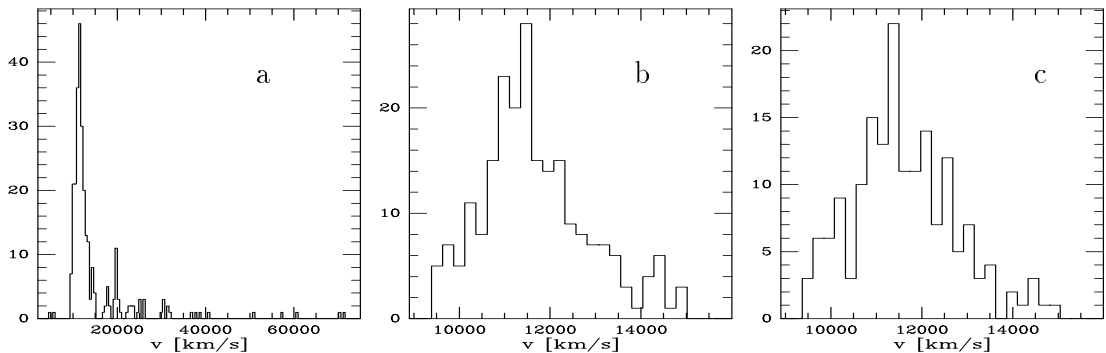

**Figure 2:** (a) The histogram of the 281 galaxies with measured velocities, (b) the entire sample of 221 cluster galaxies, and (c) the 169 cluster galaxies which form an 86% complete magnitude limited sample to $R_{23.5} = 17$. This cluster sample lies within the $2h^{-1}$ Mpc $\times 2h^{-1}$ Mpc region with CCD photometry.

### 2.2. Defining the Cluster Velocity Range

Figure 2 contains velocity histograms of the entire redshift sample and of two different cluster samples. Of the total sample of 281 galaxies, 221 lie within the velocity range of the cluster (8,500 km/s$\leq v \leq$15,500 km/s). This velocity range is defined by a 3,856 km/s gap on the low velocity end and a 1,938 km/s gap on the high velocity end; both limits are $\sim 3\sigma_v$ away from the cluster mean ($\bar{v} = 11,686 \pm 80$ km/s, $\sigma = 1,156^{+60}_{-51}$ km/s). Two hundred four galaxies with measured redshifts form an 86% complete magnitude limited sample to $R_{23.5} = 17$ within the $1° \times 1°$ ($2h^{-1}$ Mpc $\times 2h^{-1}$ Mpc) square with CCD photometry. There are 36 galaxies brighter than $R_{23.5} = 17$ without measured redshifts; Figure 3 contains a plot of their angular distribution. Out of this magnitude limited sample, 169 galaxies are within the velocity range of the cluster. Our analysis focuses primarily on this magnitude limited cluster sample.

### 2.3. Measuring the Non–cluster Background

To $R_{23.5} = 17$, there are 240 galaxies within the $1° \times 1°$ region centered on the cluster (the region with CCD photometry); 204 have measured redshifts. Thirty–five of the 204 galaxies with measured redshifts are not associated with the cluster. Assuming, conservatively, that 6 of the 36 galaxies without redshifts are background galaxies (the number is probably higher), we have an estimated background of 41 galaxies within this $\sim 1$ square degree survey region. This background is consistent with the typical field galaxy density to $R = 17$ of $\sim 45$ per square degree (Jones *et al.* 1991; Geller *et al.* 1995; Lopez–Cruz 1995).

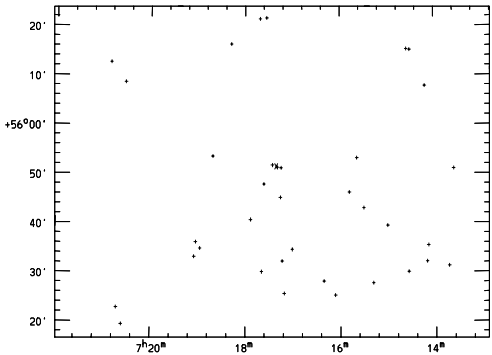

**Figure 3:** The angular distribution of the 36 galaxies brighter than $R_{23.5} = 17$ without measured redshifts. The cluster center is marked with an ×. A 2D KS test indicates this distribution is inconsistent with a homogeneous distribution (the distribution of the 204 observed galaxies) at 94.8% (96.8%) confidence.

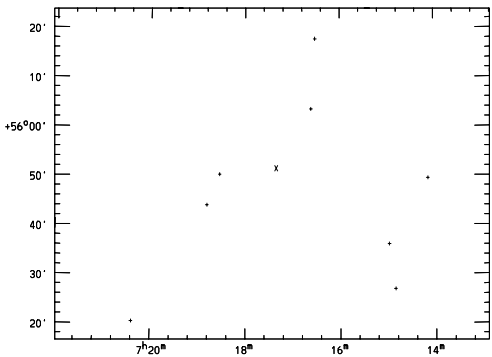

**Figure 4:** The angular distribution of the 8 galaxies in the high velocity tail ($v > 13,800$ km/s) of the cluster velocity distribution. The cluster center is marked with an ×.

*2.4. Nature of the Cluster Velocity Distribution*

The merging of subclusters is often revealed in the structure of the cluster velocity distribution (e.g. Crone & Geller 1995). Therefore, we examine the velocity distribution of Abell 576 to probe for evidence of substructure. The skewness of the velocity distribution is 0.48; the probability of a Gaussian distribution of the same mean, dispersion, and number having a skewness greater than this is $\sim 0.5\%$. The 8 galaxies in the high velocity tail ($v > 13,800$ km/s) are not clearly spatially segregated (see Figure 4). There are at least two distinct explanations: these galaxies could represent 1) a background structure at a distance of $\sim 25 h^{-1}$ Mpc beyond the cluster or 2) the remaining evidence of a subcluster merger as in Coma (Colless & Dunn 1995). In contrast to Coma, there is no clear indication of substructure in the angular distribution of cluster galaxies in A576; primarily for that reason, we do not exclude galaxies in the high velocity tail from the cluster (in §5.3 we examine the effects of these high velocity galaxies on the virial mass estimates). However, it is possible that a deeper redshift survey may uncover stronger evidence of recent dynamical activity.

The Dressler–Shectman (Dressler & Shectman 1988) statistic shows only marginal evi-

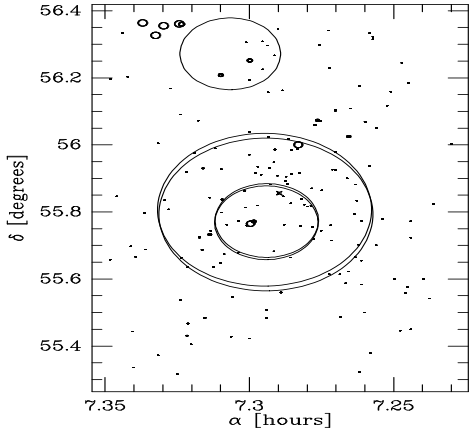

**Figure 5:** Results of the Dressler–Shectman test relying on the 5 nearest neighbors within the magnitude limited sample of 169 cluster galaxies. Monte Carlo simulations indicate that there is a 5.4% chance of obtaining a higher $\Delta_5$ with a sample this size. The ellipses have effective radii proportional to $e^\delta$, where $\delta$ is the deviation of the local mean velocity and velocity dispersion from the global values, and the cluster center is marked with an $\times$.

dence for substructure; we measure a reduced $\Delta_{10} = 1.68$ and a 23% chance of measuring a higher $\Delta_{10}$ with the sample (where $\Delta_{10}$ is the Dressler–Shectman statistic calculated using the ten nearest neighbors). We obtain similar results using the 20 nearest neighbors; with the 5 nearest neighbors there is a 5% chance of measuring a higher $\Delta_5$. The signal is dominated by 7 galaxies located $\sim 200 h^{-1}$ kpc from the cluster center (see Figure 5); these galaxies have a local mean velocity of 10,308±419 km/s and a dispersion of $546^{+501}_{-169}$ km/s (90% confidence). We regard this apparent signal as a $1.93\sigma$ noise fluctuation. Even if these 7 galaxies are part of a subcluster, it is unlikely that the subcluster affects the global cluster dynamics significantly; therefore, we include the entire cluster sample as defined in §2.2 in the following analyses.

## 3. DISTINCT GALAXY POPULATIONS

Early type galaxies preferentially frequent the high density environments typical of cluster cores (morphology–density relation; Dressler 1980). There is evidence for *kinematic* differences between the early and late type galaxies (Huchra 1985; Binggeli *et al.* 1987; Sodré *et al.* 1989; Zabludoff & Franx 1993; Colless & Dunn 1995). A direct search for these kinematic differences requires morphological typing of our sample; Melnick & Sargent (1977) classify the galaxies within the central 1.5 deg$^2$ region of A576, but publish only a radial profile of the morphological types. They conclude that the S0 fraction is high. In the absence of individual galaxy morphologies, we compare the kinematic differences between the emission line and non–emission line galaxy populations.

We divide our sample by eye into those galaxies with line emission (predominantly H$\alpha$, N1/N2, and S1/S2) and galaxies without line emission during the process of extracting redshifts. This classification by eye corresponds to minimum emission line equivalent widths which depend on the signal to noise of the sky subtracted spectra. For the brightest galaxies with the highest quality spectra, emission line equivalent widths in the H$\alpha$/N complex of 2.5–3 Å are sufficient to be classified as emission line galaxies; for the poorest spectra higher equivalent widths ($\geq 10$ Å) are required. In this division there are two important issues: 1) the Decaspec is a fiber instrument, and the input apertures subtend 2.4″ ($\sim 1.3$ kpc) diameter circles which are manually tweaked to lie on the bright, central region of the galaxy (carried out with the aid of electronic "finding charts" which display an image of the galaxy and the surrounding field), and 2) fifty of the spectra (from Hintzen *et al.* 1982 with $3''$ fibers) contain no information for $\lambda \geq 6,000$Å; the H$\alpha$ emission (typically the brightest line in our emission spectra) is outside their bandpass. Thus, although emission line spectra do imply that the galaxies contain gas and have perhaps experienced recent star formation, the lack of emission lines does not necessarily prove the absence of gas and star formation. Stated another way, we divide our galaxy sample into gas rich/gas poor subsamples, and we expect some contamination of the gas poor sample.

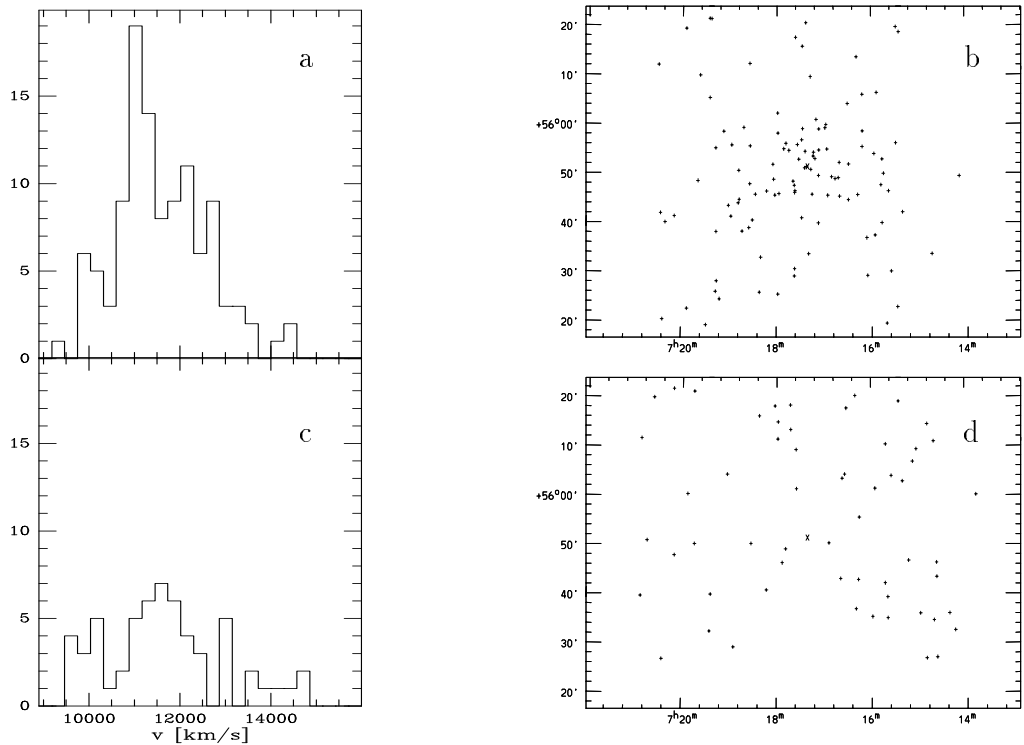

**Figure 6:** The velocity histograms and angular distributions of the emission and non–emission galaxies. On the left are the velocity histograms of the 111 non– emission line galaxies (a), and the 58 emission galaxies (c). To the right are the angular distributions of the non–emission galaxies (b) and the emission line galaxies (d). The emission line population is less clustered and has a higher velocity dispersion than the non–emission population. The cluster center is marked with an ×.

### 3.1. Kinematics and Angular Distributions

The emission/no emission samples are strikingly different in their angular and velocity distributions (see Figure 6). The 111 non–emission galaxies are well clustered around the Abell center, and the 58 emission galaxies are distributed more homogeneously. A 2 dimensional KS test clearly distinguishes between the projected non–emission sample and a homogeneous distribution (at 99.98% confidence), but fails to distinguish the projected emission sample from a homogeneous distribution (57% confidence). The emission galaxies have a larger velocity dispersion than the non–emission galaxies. The mean velocities (90% confidence; Danese, De Zotti, & di Tullio 1980) of the two samples are $\langle v_{emi} \rangle =$ $11,668 \pm 296$ km/s and $\langle v_{abs} \rangle = 11,602 \pm 160$ km/s where the subscript "abs" refers to the non–emission sample; the velocity dispersions (90% confidence) are $\sigma_{emi} = 1,297^{+239}_{-171}$ km/s and $\sigma_{abs} = 977^{+124}_{-96}$ km/s. Although a KS test fails to distinguish between the two velocity distributions, a 2 dimensional KS test (Press et al. 1992) does distinguish between the angular distributions at the 99.97% level ($D_{KS}$=0.39), and an F–test indicates that $\sigma_{emi} > \sigma_{abs}$ at the 99.4% confidence level. Given the likely contamination of the non–emission sample (§3), we stress that these observed spatial and kinematic differences are very likely real.

The skewness of the non–emission velocity distribution is 0.443 (97% significant), and the skewness of the emission sample is 0.430 (91% significant– assuming the distribution is Gaussian). We compare the angular distributions of a variety of velocity subsamples of the emission and non–emission samples as a further probe for substructure. Two dimensional KS tests fail to find differences at or above the 85% confidence level among these subsamples. The evidence for large scale recent dynamical activity in A576 is weak (see §2.4).

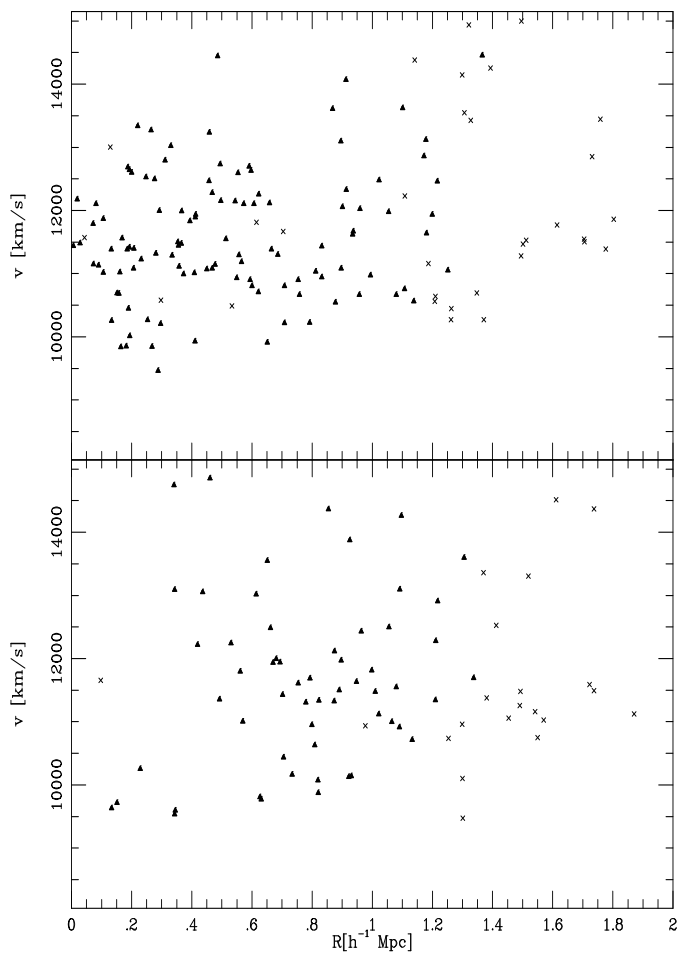

**Figure 7:** The 142 non–emission cluster galaxies (above) and the 79 emission line cluster galaxies (below) with measured redshifts plotted according to their velocity and projected distance from the X–ray centroid. Members of the magnitude limited sample are marked with filled triangles, and additional galaxies are marked with ×'s. The mean velocities and dispersions of these global samples are consistent with the means and dispersions listed in the text for the magnitude limited sample. Note that the high velocity galaxies responsible for the skewness in the velocity distribution are both emission and non–emission galaxies.

*3.2. Velocity Distribution Versus Radius*

Before studying the radial distributions of the different galaxy population, it is critical to examine the position and uncertainty of the cluster center in detail. The most accurately determined centroid in Abell 576 comes from the cluster X–ray emission observed with the *Einstein* IPC (to be discussed in §3). The cluster center is $\alpha_X$: 7:17:22.75, $\delta_X$: 55:51:22.1 (1950). The uncertainty in this centroid is dominated by systematic uncertainties inherent in centroiding X–ray sources observed with the IPC; these systematics lead to an RMS uncertainty of $\sim 15''$ (Van Speybroeck, Szczypek & Fabricant 1980). The centroids of the previously defined galaxy samples are all statistically consistent with this X–ray centroid. In particular, the distances to the centroids of the non–emission sample of 111 galaxies and the emission sample of 58 galaxies are $2'.6\pm1'.8$ and $4'.5\pm3'.2$. Therefore, we use the X–ray centroid in all radial sorting.

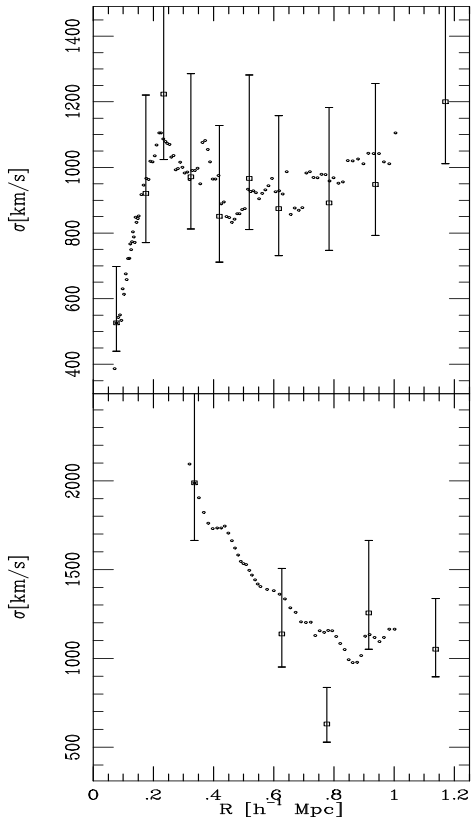

**Figure 8:** The projected velocity dispersion as a function of radius for the magnitude limited sample of 111 non–emission galaxies (above) and the 58 emission galaxies (below). The boxes with error bars (68% confidence intervals) are the dispersions of independent samples of 11 galaxies (except for the outermost bin in the emission sample which contains 14 galaxies). The remaining points represent values calculated with the sliding bin described in the text. Neighboring points are correlated. Note (1) the cool core in the non–emission sample and (2) the dispersion falling with radius in the emission sample.

Figure 7 contains velocity–radius plots for the cluster emission (142) and non–emission (79) galaxies. In both plots the members of the magnitude limited sample are marked by filled triangles, and the other galaxies are noted by ×'s. Figure 8 contains the projected velocity dispersion as a function of projected radius. These plots underscore the kinematic differences between these two galaxy populations. A 2D KS test places the probability that the velocity–radius distributions of the 58 emission and 111 non–emission galaxies are sampled from the same parent distribution at $8 \times 10^{-6}$.

At small separation from the cluster center, the paucity of emission galaxies with velocities near the cluster mean increases the dispersion. The dispersion of the 11 emission galaxies at projected separations $< 0.5h^{-1}$ Mpc is $\sigma_{<0.5} = 1,989^{+1,181}_{-518}$ km/s, compared to the global emission galaxy dispersion of $\sigma_{emi} = 1,297^{+239}_{-171}$ km/s; an F–test indicates $\sigma_{<0.5} > \sigma_{emi}$ at 98% confidence. The radial dependence of the velocity dispersion of the emission sample is reminiscent of the behavior of the spherical infall model (Regös & Geller 1989); the complexity of the radial distribution of velocities (Figure 7) is consistent with the expected deviations from spherical infall (van Haarlem 1992), but is probably also contaminated by distant ($\sim 20$ Mpc) galaxies projected along the line–of–sight. The dispersions of the emission and non–emission samples converge at a value of $\sim$1,000 km/s outside the core ($> 0.5h^{-1}$ Mpc); the 51 non–emission galaxies have $\sigma_{abs} = 992^{+198}_{-138}$ km/s and the 47 emission galaxies have $\sigma_{emi} = 1,106^{+232}_{-159}$ km/s at 90% confidence.

Figure 7 and Figure 8 also indicate that the velocity dispersion of the non–emission sample drops significantly in the core. In §4 we discuss this cool core in the galaxy population.

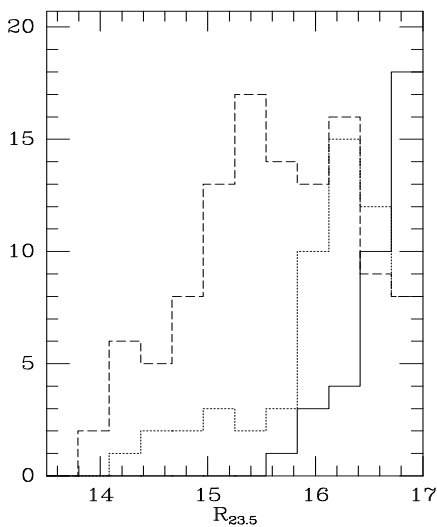

**Figure 9:** Histograms of the $R$ band magnitude distributions of three galaxy samples brighter than $R_{23.5} = 17$: the 58 emission line galaxies (short dashed), the 111 non–emission line galaxies (long dashed), and the 36 unobserved galaxies (solid line). The median magnitudes of the emission, non–emission and unobserved samples are 15.7, 16.2, and 16.7.

### 3.3. Luminosity Distribution

The non–emission galaxy sample is significantly brighter than the emission sample (see Figure 9). A KS test places the probability that the emission and non–emission sample are drawn from the same parent population at $2 \times 10^{-6}$. The median (mean) magnitudes of the 58 emission galaxies and the 111 non–emission galaxies are 16.2 (16.1) and 15.7 (15.6).

A similar magnitude offset in the luminosity functions of emission and non–emission galaxies has also been observed in the Las Campanas Redshift Survey (Lin *et al.* 1995). The best fit Schechter functions to the Gunn–$r$ luminosity functions are $M_* = -20.03 \pm 0.03$, $\alpha = -0.9 \pm 0.1$ for the emission galaxies and $M_* = -20.22 \pm 0.02$, $\alpha = -0.3 \pm 0.1$ for the non–emission galaxies. Our limited sample precludes a meaningful comparison of the best fit Schechter parameters for the emission and non–emission luminosity functions; however, in §5.1 we discuss the cluster luminosity function to $R_{23.5} = 19$ in detail.

The $R$ band offset between the magnitude distributions of the emission and non–emission galaxies probably reflects the $\Delta(B - R) \sim 1$ mag color difference between early and late–type galaxies. The $B$ band luminosity function of the early type galaxies in Virgo is slightly brighter than that of the late type galaxies (Binggeli, Sandage & Tammann 1988 and refs. therein); given the color difference between these types, the magnitude offset in the $R$ band would be larger. To the extent that the emission/non–emission classification runs parallel to the late/early type classification, the Virgo observations suggest that the observed magnitude offset in A576 is caused by a color difference between the two populations.

### 3.4. Discussion

The $R_{23.5} < 17$ magnitude limited sample provides strong evidence that the 58 emission and 111 non–emission galaxies with redshifts are distinct populations. The angular distributions (99.98% confidence) and the velocity dispersions (99.4% confidence) are distinct (§3.1). In addition, the emission line galaxies have a very different $R$ band magnitude distribution than the non–emission galaxies ($10^{-6}$ chance of consistency, §3.3). The behavior of the velocity distribution as a function of radius for the two populations is markedly different ($10^{-5}$ chance of consistency; see Figure 7, §3.2); the non–emission galaxies contain a cool core (99.8% confidence), and the emission galaxies have a velocity dispersion which peaks in the core (98% confidence).

These two spectroscopically defined subsamples probably parallel the early/late type morphological classification scheme. The data in A576 then provide strong confirmation that the observed angular segregation which has long been known as the morphology–density relation (Dressler 1980) is coupled to kinematic differences (Huchra 1985; Binggeli *et al.* 1987; Sodré *et al.* 1989; Zabludoff & Franx 1993, Colless & Dunn 1995).

On the $2h^{-1}$ Mpc scale of the CCD photometry the angular distribution of the emission population is indistinguishable from a homogeneous distribution (§3.1). Three competing explanations are that the emission population could be (1) a virialized population with higher dispersion and a much shallower density profile, (2) a background population residing at large ($\sim 10$ Mpc) distances from the cluster, or (3) a non–virialized near–core population on its initial infall into the cluster. If the emission population is virialized, then its environment is similar to that of the non–emission population, and the differences in gas content are a mystery. Although the angular distribution of the sample (on this scale) is consistent with either of the non–virialized hypotheses, the kinematic evidence (Figure 7, Figure 8) favors the near–core population on initial infall (at 98% confidence). Specifically, the emission galaxies projected near the cluster core have velocities which appear to avoid the mean cluster velocity; this distribution drives the higher velocity dispersion of the emission sample at small separation from the cluster center. In the background scenario the dispersion of the emission population should be independent of radius; the observed increase in the dispersion with decreasing radius would be a mere coincidence. For the initial infall scenario the rising dispersion in the core is expected (Regös & Geller 1989, van Haarlem 1992).

If the emission population is a near–core population on initial infall, then these observations in A576 are consistent with models in which star formation is suppressed in the cluster core (Dressler & Gunn 1982, 1983; Bothun & Dressler 1986). Additionally, although our emission/non–emission classification is a blunt, binary scheme, we note that emission spectra are common in our field populations; it is the lack of star formation in the core population (non–emission galaxies) which is remarkable, not the presence of star formation in the near–core population (emission galaxies).

## 4. THE COOL CORE

Although A576 shows no obvious evidence for a recent merger, it is not a simple system; in this section we show that the core of the cluster is dominated by a galaxy population and an ICM with temperatures significantly less than the global values. We discuss two heuristic models as a basis for a consistent dynamical picture of A576.

### 4.1. Core Galaxy Population with Low Velocity Dispersion

Figure 7 and Figure 8 show that A576 contains a low velocity dispersion population in the non–emission sample; the 10 galaxies nearest the cluster center (within $132h^{-1}$ kpc) have a velocity dispersion of $\sigma_{core} = 387^{+250}_{-105}$ km/s (uncertainties are 90% confidence limits). This value is much smaller than (2.5×) the global dispersion of the non–emission population ($\sigma_{abs} = 977^{+124}_{-96}$ km/s); an F–test indicates that $\sigma_{abs} > \sigma_{core}$ at 99.8% confidence. Under the assumption that the cool core galaxies are contained within a volume with radius equal to the projected radius, the galaxy density within the core is $\sim 10^3 h^3$ Mpc$^{-3}$, corresponding to a galaxy overdensity of $\sim 3 \times 10^4$ (Lin et al. 1995).

In an attempt to understand whether this phenomenon can be explained as mass segregation, we examine the photometry of the low dispersion population in more detail. The ten galaxies which form the cool core span a range in isophotal magnitude ($14.3 \leq R_{23.5} \leq 16.4$) and central surface brightness ($16.7 \leq \mu_R \leq 18.2$) comparable to the non–emission population outside the core ($14.0 \leq R_{23.5} \leq 17.0$ and $16.5 \leq \mu_R \leq 19.3$); a KS test fails to distinguish between the magnitude or surface brightness distributions of the two populations. Thus, if these galaxies are a more massive population and have cooled through relaxation effects, isophotal $R$ band light is not a good indicator of individual galaxy mass. Alternatively, under the assumption of a correlation between mass and light, relaxation effects can be ruled out as the cause of the low dispersion galaxy population. We emphasize that the existence of the fundamental plane for early type galaxies and the bulges of lenticulars (e.g. Jøergensen, Franx, & Kjærgaard 1993) favors the latter intepretation.

### 4.2. ICM Temperature

There is also clear evidence for a cool core in the ICM. Unfortunately, there is no angularly resolved spectrum of A576. The *Einstein* Solid State Spectrometer (SSS) observation of the central $202h^{-1}$ kpc ($6'$ diameter) region of A576 yields a gas temperature of $1.6^{+0.4}_{-0.3}$ keV (90% confidence) with roughly half solar abundances (Rothenflug et al. 1984). Recent reductions of the *Einstein* Monitor Proportional Counter (MPC) data (David et al. 1993) reveal that the cluster emission over a $1.45h^{-1}$ Mpc scale ($43'$ FWHM; Grindlay et al. 1980) has a best fit single temperature model with $kT = 4.3^{+0.5}_{-0.4}$ keV (90% confidence), consistent with earlier observations by HEAO 1 A–2 (Rothenflug et al. 1984).

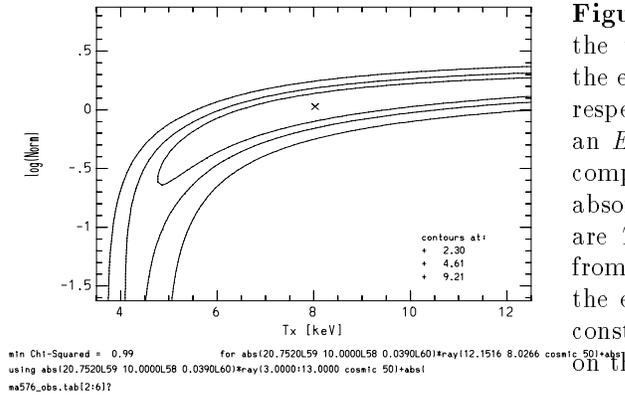

**Figure 10:** A contour plot of $\chi^2$ as a function of $T_X$, the temperature of the hot component, and $Norm$, the emission normalization of the cool component with respect to the hot component. The $\chi^2$ is calculated for an *Einstein* MPC spectrum (David 1995) fit to a two component Raymond–Smith spectrum with galactic absorption. The best fit values (marked with an $\times$) are $T_X = 8$ keV and $Norm = 1.0$, equal emission from the hot and cold component. Note that unless the emission fraction of the cold gas is independently constrained, the MPC spectrum places no upper limit on the temperature of the hot component.

The global temperatures for A576 assume emission from an isothermal gas. However, the observations with the SSS indicate that at least 15% of the emission is coming from a cool component (Rothenflug et al. 1984). We examine fits of a two temperature model to the reduced MPC data (David 1995). We fix the low temperature component to $kT = 1.6$ keV and allow the temperature of the hot component to vary. As the emission fraction of the low $T$ component increases, the best fit temperature of the hot component also rises. With the emission fraction as a free parameter (the SSS observation provides no information about the amount of cool gas outside its $200h^{-1}$ kpc field of view), the best fit emission fraction is 50% and the temperature of the hot component is 8 keV. A $\chi^2$ map (5 constraints and 3 free parameters) indicates that unless the emission fraction of the cold component is known, the constraint on the temperature of the hot component is $> 4$ keV (at 90% confidence); the available MPC data place no upper limit on the temperature of the hot component. Interestingly, combining the velocity dispersion of the non–emission sample (977 km/s) with the $\sigma$–$T$ relation implies a gas temperature $T \sim 6$ keV (Lubin & Bahcall 1993).

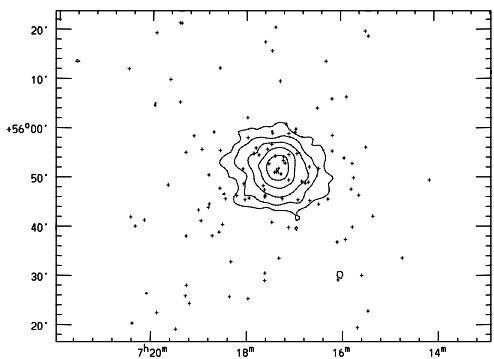

**Figure 11:** A contour plot of the *Einstein* IPC image of the X–ray emission from A576 with the non–emission cluster galaxy sample overplotted. The 5 contours are offset in equal logarithmic intervals between the peak surface brightness ($I_0 = 3.8 \times 10^{-13}$ ergs/s/cm$^2$/arcmin$^2$) and 6% of that value. The X–ray centroid is marked with an $\times$.

## *4.3. X–ray Luminosity*

The X–ray spectral observations indicate that the gas in A576 has a cold core and a hotter global temperature. This temperature variation could in principle affect luminosity calculations. In the absence of a temperature profile we estimate the cluster X–ray luminosity using the *Einstein* IPC observation and a range of global gas temperatures. Two *Einstein* archival Imaging Proportional Counter (IPC) observations of A576 are available (see White & Silk 1980 for discussion). We reduce the 10,377 s image in the standard way (Mohr, Fabricant & Geller 1993) to a Gaussian smoothed resolution of $2.4'$ FWHM. The reduced 0.3–3.5 keV image contains $\sim$7,400 cluster photons. If the emission comes predominantly from a (1.6, 4.3, 9.0 keV) Raymond–Smith spectrum with half solar abundances, the 0.3–3.5 keV luminosity is $(4.3, 4.5, 4.6) \times 10^{43} h^{-2}$ ergs/s (calculations done with PROS software developed at SAO), in reasonable agreement with the luminosity in White & Silk (1980). The bolometric luminosity is uncertain by $\sim$50% because of the unknown temperature variation; we calculate that the bolometric luminosity is $(6.4, 8.2, 12) \times 10^{43} h^{-2}$ ergs/s for the same range of gas temperature. The bolometric luminosity for the $kT = 4.3$ keV gas is within 11% of the value obtained by David et al. (1993) with the MPC.

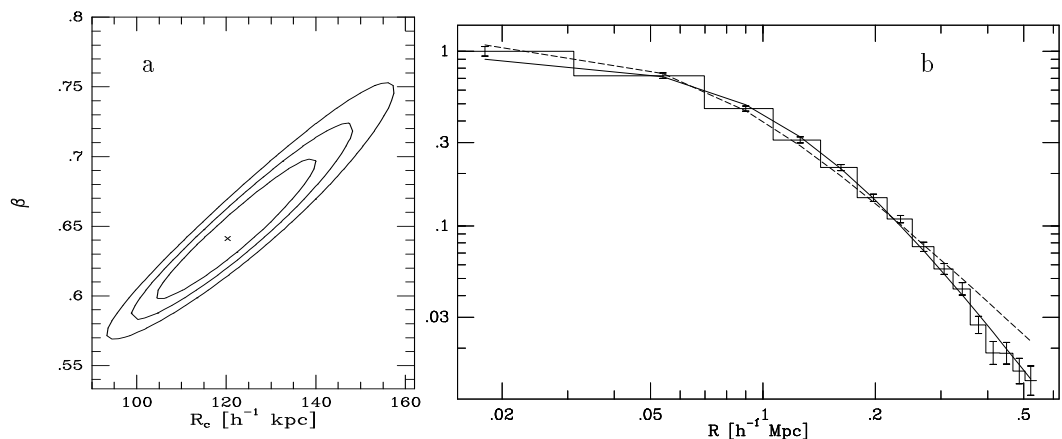

**Figure 12:** Fit to the radial profile of the X–ray emission. Figure b is the radial profile of the cluster X–ray emission (histogram with error bars) and the best fit $\beta$ model (solid line). The dashed line is the best fit model with fixed $\beta = 0.5$ ($\chi_\nu^2 = 7.7$). Figure a is a contour map of $\chi^2$ as a function of $\beta$ and $R_c$. The $\Delta \chi^2$ values correspond to $1\sigma$, $2\sigma$ and $3\sigma$ confidence limits. The best fit $\chi_\nu^2 = 1.51$ (11% chance of consistency between the fit and the data), and the best fit parameters (marked with an $\times$) are $I_0 = 3.8 \times 10^{-13}$ ergs/s/cm$^2$/arcmin$^2$, $R_c = 120 h^{-1}$ kpc, and $\beta = 0.64$. A power law fit to the 9 outermost points ($0.25 \leq Rh \leq 0.5$ Mpc) yields a $\beta = 0.65$.

### 4.4. ICM Distribution

We azimuthally average the 0.3–3.5 keV X–ray emission around the X–ray centroid and fit it to a $\beta$ model (Cavaliere & Fusco–Femiano 1978).

$$I(R) = I_0 \left( 1 + \left( \frac{R}{R_c} \right)^2 \right)^{-3\beta + \frac{1}{2}} \qquad (4.1)$$

We use the unsmoothed IPC image with resolution FWHM=$50 h^{-1}$ kpc to calculate the radial profile (see Mohr et al. 1995). The best fit model has a $\chi^2 = 1.51$ (corresponding to 11% probability of consistency between the fit and the data; see Figure 12); the fit parameters are $I_0 = 3.8 \times 10^{-13}$ ergs/s/cm$^2$/arcmin$^2$, $R_c = 120 h^{-1}$ kpc, and $\beta = 0.64$. The correlation of $\beta$ with $R_c$ is clearly visible in the $\chi^2$ contour map (Figure 12); the best fit value is marked with an $\times$, and the surrounding contours correspond to 1, 2, and $3\sigma$ confidence levels.

White & Silk (1980) fit the radial emission profile with a Hubble profile (fixed $\beta = 0.5$), and Jones & Forman (1984) fit with the standard $\beta$ model. The best fit $\beta$ parameters fit by Jones & Forman (1984) lie well outside our statistical error range; their fit has a core radius range of $R_c$ =45–70$h^{-1}$ kpc and radial fall–off range of $\beta$ =0.47–0.52. The dashed line in Figure 12b is the best fit model with fixed $\beta = 0.5$. The IPC resolution does not affect the emission profile in Figure 12 because the measured core is $\sim 5\times$ the resolution FWHM. We suspect that the discrepancy between our result and the Jones & Forman values comes from using a somewhat different radial range. Jones & Forman (1984) extend their profiles as far as $32'$ (masking the ribs to avoid contamination), and we truncate our profile at $16'$, well within the ribs where the internal detector background is lowest. Our measured core radius still places A576 within the XD class defined by Jones & Forman (1984).

### 4.5. Central Cooling Time

Next we investigate the central cooling time of the gas using the approach described by David *et al.* (1990). We are hampered in this calculation by the unknown temperature variation in the cluster. However, using the range of bolometric luminosities calculated assuming isothermal emission, we estimate that the central density is $3.5 \times 10^{-3} h^{1/2}$ cm$^{-3}$, consistent with the value found by Rothenflug *et al.* (1984) and Jones & Forman (1984). The central density and the central temperature yield a central cooling time of $1 \times 10^{10}$ yrs, consistent with previous results (Rothenflug *et al.* 1984; Edge, Stewart & Fabian 1992).

The cooling time indicates that this cluster is marginally unstable to cooling. Cooling flows are usually centered on cD galaxies (Jones & Forman 1984), but in A576 there is no large central galaxy. However, there are two bright early type galaxies within an arcmin of the X–ray centroid, and they may share a common envelope (Rothenflug *et al.* 1984). Unfortunately, there is also an 8th magnitude star within an arcmin of these galaxies. The scattered light from the star makes it difficult to measure accurate magnitudes or to determine whether the galaxies share a common envelope; however, the relative line of sight velocity of 740 km/s makes the shared envelope hypothesis improbable. Nevertheless, a weak cooling flow centered on one of the two central ellipticals is not ruled out by the data.

### 4.6. Gas Cooling Flow and Galaxy Mass Segregation

Both the X–ray emitting gas and the non–emission galaxies indicate a cool core population embedded in a hotter global population. We suggest two models which might account for the data and then discuss ways of discriminating between them. In broad terms either physical processes which are more efficient in the core have led to the emergence of cool populations out of initially homogeneous and hot distributions, or the gas and galaxy core of A576 has always been cooler than the global populations. Of course, the galaxies and gas need not be described by the same model.

In the first model the initial collapse and violent relaxation of the cluster leaves the galaxy and gas populations with roughly uniform energy per unit mass. Following the initial collapse, the galaxy population evolves toward energy equipartition (e.g. Chandrasekhar 1942; Binney 1977; Frenk *et al.* 1995), and the densest gas cools significantly through radiative losses (e.g. Lea *et al.* 1973; Edge *et al.* 1992). The central cooling time in the gas is $10^{10}$ yrs. We obtain an estimate of the galaxy relaxation timescale using the median relaxation time (Spitzer & Hart 1971; Binney & Tremaine 1987; den Hartog 1995). Applied to the central 1.0 Mpc of a cluster of enclosed mass $2.8 \times 10^{14} M_\odot$ (cluster mass discussed at length in §5.3) with galaxies of typical mass $m_{gal} = 10^{11} M_\odot$, this estimate implies a relaxation timescale of the order of $1.2 \times 10^{10}$ yrs (this relaxation time scales as $1/m_{gal}$). This scenario requires cooling of the core gas and galaxies for a significant fraction of the Hubble time.

The cooling times for the gas and the galaxies are marginally consistent with this scenario as long as the cluster has evolved in relative isolation for a good fraction of the Hubble time. Because different processes are responsible for the cooling in the galaxies and the gas, it is unlikely that the ratio of the energy per unit mass (e.g. Sarazin 1988; Lubin & Bahcall 1993; den Hartog 1995) in the galaxies to that in the gas ($\beta = \sigma^2 \mu m_p / k_B T_X$) would be maintained at its original value in the cooling region of the cluster. The central value is $\beta = 0.6^{+1}_{-0.3}$ and the value outside the core is $\beta = 1.5^{+0.4}_{-0.3}$ (4 keV/$T_x$). These values are too poorly constrained to be useful; a larger redshift sample in the core and a gas temperature profile would allow a more accurate $\beta$ comparison.

Additional spectral observations of the X–ray emitting gas would test the gas cooling flow hypothesis. In particular, with a gas temperature profile one can determine whether the cool gas core is confined (as expected in this scenario) to the region where the cooling time is less than a Hubble time. The cooling time reaches $2 \times 10^{10}$ yrs at a radius of $\sim 55 h^{-1}$ kpc; angularly unresolved observations with the SSS over a region roughly twice this size demonstrate evidence for cool gas (Rothenflug *et al.* 1984). Angularly resolved observations with evidence for cool gas at larger radii would serve as evidence that radiative losses alone can not account for the cool gas core.

### 4.7. Merger and Survival of a Subcluster

An alternative model involves the merger and survival of a subcluster. In this model the core gas and galaxies reach their present cool temperatures through the gravitational collapse of a structure less massive than A576. The present epoch configuration comes about either through a merger of the subcluster with the mostly collapsed A576, or through the formation of A576 around the previously collapsed subcluster. In the former case, the subcluster must survive the tidal stresses of merging with a massive cluster (Merritt 1984); in the latter case, the subcluster must survive the infall and heating of gas and galaxies as the entire cluster forms. The central issue in this scenario is to understand under what conditions (if any) the subcluster can survive the relaxation processes which tend to erase all but the highest contrast substructure– galaxies (White & Rees 1978).

We consider this scenario for several reasons. It is theoretically appealing because of the abundance of evidence for cluster substructure (Geller & Beers 1981; Dressler & Shectman 1988; Forman & Jones 1992; Mohr *et al.* 1995), and because of an apparent consistency with the popular bottom–up models of structure formation. Observationally, several issues lead us to question the cooling flow and mass segregation scenario; these issues include: (1) the evidence for a cooling flow (typically the presence of a cD galaxy and a significant X–ray emission excess; Forman & Jones, 1984) is weak, (2) there is weak evidence of cool gas at roughly twice the cluster cooling radius, and (3) the cool galaxy core is composed of bright *and* faint galaxies. None of these observational "complications" are problems for the second scenario. Unlike many other clusters there is no compelling evidence for recent dynamical activity in the galaxy angular distribution, ICM distribution, or galaxy velocity distribution; these observations are consistent with the merger hypothesis only if the merger occurred well before the present epoch.

However, the subcluster survival issue poses a serious challenge. The curvature in the cluster potential truncates a galaxy orbiting near the cluster core to a radius $r_T \sim R_c \sigma_g / \sigma_{cl}$ where $R_c$ is the cluster core radius, and $\sigma_g$ ($\sigma_{cl}$) is the galaxy (cluster) velocity dispersion (Merritt 1984). This same process would effectively truncate a merging subcluster. A truncation radius of $200h^{-1}$ kpc requires a $\sim 500h^{-1}$ kpc core radius for the cluster potential (taking Merritt's $\alpha = 2$, an isothermal mass distribution in the subcluster); the gas core radius ($120h^{-1}$ kpc; see Figure 12) and the non–emission galaxy core radius ($< 500h^{-1}$ kpc unless the galaxies fall off more steeply than $r^{-3.4}$; see §5.2) are both smaller than this. However, the truncation radius depends sensitively on the position of the subcluster within the cluster. In particular, the values quoted are the most extreme, holding if the subcluster is in a circular orbit at one cluster core radius. A more realistic model would entail generally radial infall, lessening the effects of the cluster tidal field.

Radial infall would pose additional problems for the subcluster gas. In a case where the subcluster infall velocity is supersonic, shocks form and thermalize the kinetic infall energy of the subcluster gas. N–body and hydrodynamics simulations indicate that these relaxation processes are efficient; substructure is generally erased on the order of a crossing time after the merger (e.g. Evrard 1990; Evrard *et al.* 1993).

If the cluster collapses with the subcluster already at its core the tidal truncation problem is eliminated. However, the core of cool gas must remain insulated from the hot gas to survive to the present epoch. We estimate the thermal conduction timescale as $t_{cond} = n_e l_T^2 k_B / \kappa$ (Sarazin 1988) where $n_e$ is the electron density, $l_T \equiv |T_e/\nabla T_e|$ is the length scale of the electron temperature $T_e$ variation, $k_B$ is the Boltzmann constant and $\kappa$ is the thermal conductivity. Using the Spitzer (1956) value for $\kappa$ (appropriate for unsaturated conduction in an ionized hydrogen plasma), and approximating $l_T \sim r_{out} T_{hot}/(T_{hot} - T_{cold})$, we calculate a conduction time of

$$t_{cond} \sim 3.4 \times 10^9 h^{-3/2} \mathrm{yrs} \left( \frac{r_{out}}{200 \text{ kpc}} \right)^2 \quad (4.2)$$

where we have taken $n_e = 2 \times 10^{-3}$ cm$^{-3}$, $T_{hot} = 4$ keV, and $T_{cold} = 1.6$ keV. If magnetic fields are present then the conduction is suppressed. For the case of magnetic fields tangled on scales smaller than $l_T$, the conduction timescale is increased by about a factor of 3 (Sarazin 1988). These estimates indicate that it may be possible for a cool gas core to survive embedded in hot gas for $\sim 10^{10}$ yrs. More detailed calculations are required for verification.

In addition, the low velocity dispersion of the subcluster galaxy population must be stable over times comparable to the Hubble time. Two body interactions with (massive) high dispersion galaxies will tend to heat the low dispersion galaxies. The timescale for this heating is related to the median relaxation timescale; there is a correction mostly due to the lower kinetic energy in the cooler population. Using the form of the diffusion coefficient listed in Binney & Tremaine (1987) we calculate the timescale correction for a population of galaxies with $\sigma_c = 387$ km/s embedded in a population with $\sigma_h = 977$ km/s. The correction is $\sim 2.5(\sigma_c/\sigma_h)^2 = 0.4$, implying that the timescale for orbital decay is $\sim 5 \times 10^9$ yrs. Thus, depending on the timescale of cluster formation for A576, some heating of the core population may be expected in this scenario.

### 4.8. Subcluster Remnant Versus Mass Segregation

Although there are problems for both of the heuristic models, we emphasize that the data do not rule either one out. In addition, the relaxation timescales which provide hurdles for these models are sensitive to the mass scale of the populations (galaxies, dark matter) responsible for the relaxation. These mass scales are not well determined observationally.

The mass segregation model is weakened primarily by the broad range of $R$ band luminosities in the galaxies which make up the cool core (unless, of course, galaxy luminosity and mass are uncorrelated). Yet observations of other clusters indicate the presence of cool cores (Cowie & Hu 1986; Bothun & Schombert 1990; Merrifield & Kent 1991, den Hartog 1995); some of these populations are associated with brightest cluster galaxies, but others– A576, in particular– are not. These cool populations are presumably not exclusively the brightest (most massive) galaxies because den Hartog (1995) finds only a weak correlation between an "inverted" dispersion profile and the presence of luminosity segregation. Theoretical arguments by Merritt (1985) indicate that the galaxy capture rate by brightest cluster members is very low in rich clusters; it scales as $\sigma_{cl}^{-7}$. Thus, perhaps even for the cool populations bound to brightest cluster galaxies, a more appealing model would be the merging and (partial) survival of a subcluster or the formation of a cluster around a previously formed subcluster.

We note in passing that another explanation for the cold core is the superposition along the line of sight of two spatially distinct clumps of galaxies. This hypothesis is highly unlikely primarily because (1) the mean velocities of the two systems are indistinguishable (implying they are gravitationally bound and near turn–around), (2) the mean angular positions of the X–ray emitting gas and the galaxies for the two systems are indistinguishable (implying precise alignment), and (3) no high velocity galaxies from the more massive system are detected at small projected radius even though the surface density would be expected to peak at $R = 0$ (for a magnitude limited sample this requires an incredible coincidence).

## 5. GLOBAL CLUSTER CHARACTERISTICS

Along with the evidence for distinct galaxy populations and a cool core in the galaxies and the ICM, the data in A576 are ideal for investigating global descriptors. Below we evaluate the luminosity function and calculate the range of radial models consistent with the galaxy distribution. We use the data to compare the equilibrium mass estimates from the galaxies and the X-ray emitting gas, and consider explanations for the significant differences. Then we use the mass range to place limits on the $R$ band mass-to-light ratio and on the cluster baryon fraction.

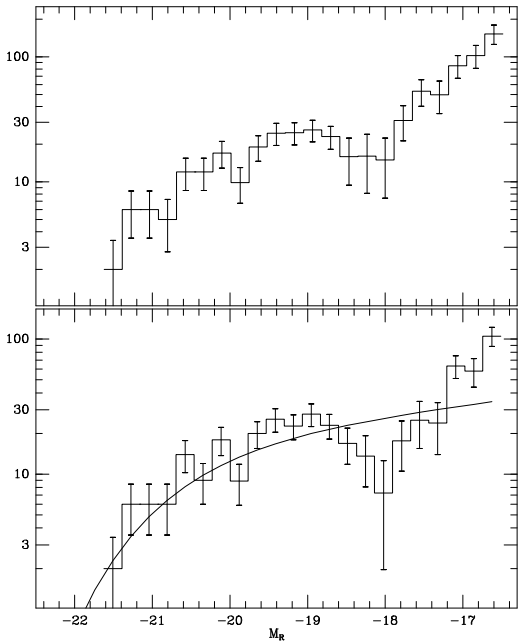

Figure 13: Cluster luminosity function. The upper figure contains the magnitude distribution of all known cluster galaxies brighter than $M_R = -18.5$ (with a completeness correction applied to each bin), and all galaxies with absolute magnitudes between $M_R = -18.5$ and $M_R = -16.5$. The lower figure is the magnitude distribution and best fit Schechter function for the same sample after statistical background subtraction in the range $M_R = -18.5$ to $M_R = -16.5$.

### 5.1. Luminosity Function

Although the data indicate that there are at least two distinct galaxy populations with very different $R$ band luminosity characteristics in A576, the magnitude limited sample of galaxies is too small for a meaningful comparison of the best fit Schechter parameters for the two populations. We examine the combined luminosity function for $R_{23.5} \leq 19$, two magnitudes fainter than the limit of our complete redshift sample. Specifically, we fit to the Schechter function (Schechter 1976, Dressler 1978) of the form

$$\left.\frac{dn}{dM}\right|_M \propto 10^{0.4(1+\alpha)(M_*-M)} \exp\left[-10^{0.4(M_*-M)}\right] \qquad (5.1)$$

where $M_*$ is the characteristic absolute magnitude and $\alpha$ is the faint end slope. The best fit parameters minimize the $\chi^2$ between the analytic form and the binned galaxy sample. Although the data indicate that the luminosity distribution varies with galaxy population, we explore the composite distribution to fainter magnitudes to provide constraints on the total cluster $R$ band light. Within the $1° \times 1°$ region with CCD photometry we identify a sample of 977 galaxies brighter than $R_{23.5} = 19$. We use only the confirmed cluster members brighter than $R_{23.5} = 17$ (making the appropriate completeness correction in each magnitude bin) and make a statistical background subtraction for fainter magnitudes. We use the shape of the background measured by Lopez–Cruz (1995) from deep $R$–band CCD photometry in the field; this background is within $\sim 30\%$ of that obtained by Jones et al. (1991) over the range of interest. In estimating the uncertainties in each bin of the luminosity function, we include a 10% uncertainty in the background correction as well as the Poisson contribution.

Because we have only $R$ band photometry, we apply the $K$–correction for a galaxy with $B - R = 1.5$ and no evolution; the appropriate value is 0.03 mag (McLeod 1995). The galactic extinction, calculated from the neutral hydrogen column density, is $A_R = 0.13 \pm 0.04$ mag (Burstein & Heiles 1978, Rieke & Lebofsky 1985). Combining these corrections, the apparent magnitude limit $R_{23.5} = 19$ corresponds to an absolute magnitude limit $M_{lim} = -16.48 + 5 \log h$. We apply these corrections to the absolute magnitudes; for comparison with other work it is important to note that the faint isophotal limit of $R = 23.5$ mag/arcsec$^2$ is not corrected for galactic extinction or cosmological dimming.

The best fit Schechter function has the parameters $M_* = -20.8^{+0.6}_{-0.3}$ and $\alpha = -1.18^{+0.13}_{-0.11}$ (see Figure 13). The reduced $\chi^2 = 3.0$ with 22 degrees of freedom, formally implying a vanishing probability of consistency between the fit and the data. The parameter ranges correspond to a 90% confidence interval, calculated in a Monte Carlo fashion and accounting only for the statistical uncertainties. This faint end slope is inconsistent with the slope of the Gunn–$r$ band luminosity function for a sample of 18,678 galaxies observed in the LCRS ($\alpha = -0.70 \pm 0.05$ to a limiting magnitude of $M_r = -17.5$; Lin et al. 1995), but consistent with the slope in the CfA Redshift Survey Zwicky magnitude luminosity function for galaxies beyond $cz = 2,500$ km/s ($\alpha = -1.0 \pm 0.2$ to a limiting magnitude $M_Z = -13$; Marzke, Huchra & Geller 1994). The Schechter parameters in A576 lie within the broad range measured in other Abell clusters (Lugger 1986, Oegerle, Hoessel & Jewison, 1987).

Finally, the magnitude distribution steepens for $M_R \geq -18$ even after statistical background subtraction. The steepening can not be explained by incorrect background normalization; removing the effect would require a different background shape. This observed steepening is potentially related to the faint–end excess noted in the CfA Redshift Survey (Marzke et al. 1994) within the range $-18.5 \leq M_R \leq -14.5$; however, it should be noted that the background correction is large in the faint bins of the A576 sample.

### 5.2. Radial Galaxy Distribution

We examine the radial distributions of the galaxy samples by fitting the projected distributions to the function

$$\nu(R) = \nu_0 \left(1 + \left(\frac{R}{R_c}\right)^2\right)^{1/2 - \alpha/2} \tag{5.2}$$

which allows for a core radius $R_c$ and a variable radial fall–off $\alpha$; this model is essentially the $\beta$ model (Cavaliere & Fusco–Femiano 1978). The best fit parameters minimize the $\chi^2$ fit of the binned galaxy samples (to a projected radius of $1h^{-1}$ Mpc) and the model (see Figure 14). The uncertainties in the measured parameters are dominated by sampling noise; the contribution associated with the centroiding uncertainty is negligible because of the accuracy of the X–ray centroid (this analysis assumes that the galaxy distribution is generally spherical and centered on the minimum of the cluster potential).

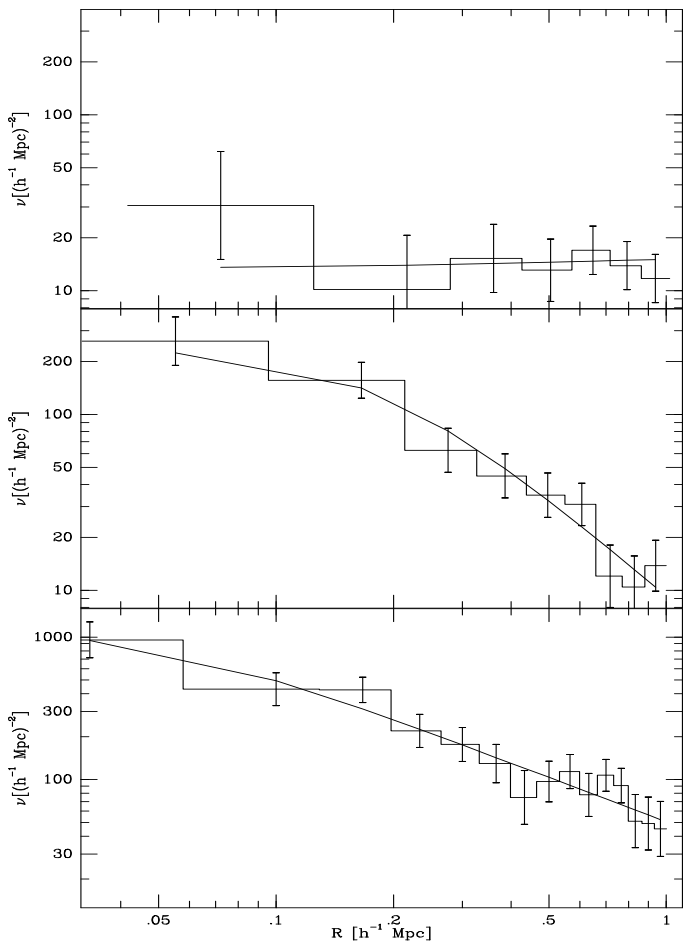

**Figure 14:** Observed and best fit radial profiles for three different ensembles of cluster galaxies: the 58 emission line galaxies (top), the 111 non–emission galaxies (middle), and the background corrected sample to $R_{23.5} = 19$ (bottom). The fit functions are of the form in Equation 5.2.

Given the correlation between the parameters $R_c$ and $\alpha$ (larger core radius favors steeper radial falloff), the constraints on the radial falloff with a sample of $\sim 100$ galaxies are weak (see Figure 15). Consider the non–emission and emission samples as an example. A 2D KS test distinguishes between the two samples with high confidence (§3.1), and the radial profiles appear strikingly different (see Figure 14). However, both samples are consistent with a radial falloff of $r^{-2}$ ($\chi^2$ probability of consistency $\geq 20\%$). Attempting to increase the sample by going fainter (where there are few measured redshifts) leads to uncertainties associated with the poorly known background correction. For the entire sample of galaxies to $R_{23.5} = 19$ we subtract the expected non–cluster background between $R_{23.5} = 17$ and 19 before fitting. The best fit background subtracted radial falloff is $\sim r^{-2.0}$.

Figure 15 contains a map of the $2\sigma$ $\chi^2$ confidence contours as a function of the parameters $\alpha$ and $R_c$ for the emission, non–emission and deep samples. The best fit core radii (and 90% confidence intervals) with the deprojected radial falloff $r^{-3}$ (e.g. Dressler 1978; Kent & Gunn 1982; Binggeli *et al.* 1987) for the emission, non–emission, emission + non–emission, and deep samples are $> 505 h^{-1}$ kpc, $180^{+136}_{-67} h^{-1}$ kpc, $311^{+275}_{-127} h^{-1}$ kpc, and $266^{+220}_{-108} h^{-1}$ kpc. The core radius for the combined non–emission + emission sample is consistent with the observed range in other clusters (Bahcall 1975; Dressler 1978).

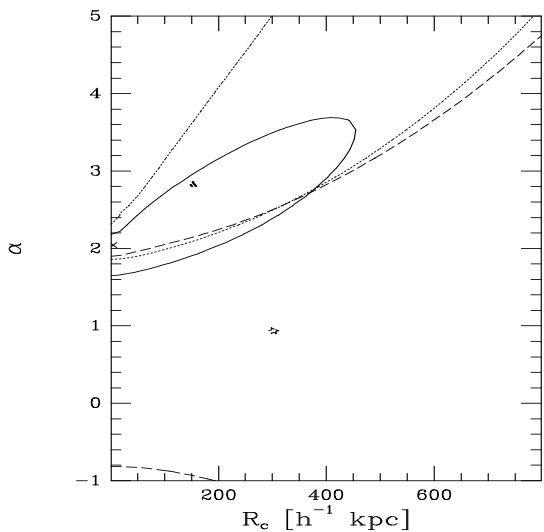

**Figure 15:** The $2\sigma$ $\chi^2$ contours of the radial galaxy distributions in $\alpha$–$R_c$ space (see Equation 5.2). The $2\sigma$ contour (solid line) for the sample to $R_{23.5} = 19$ is the best constrained; the best fit value is marked with an $\times$ (near the $\alpha$ axis). The contour for the 111 non–emission (58 emission) galaxies is the dotted line (dashed line), and the best fit value is the filled triangle (star). The $2\sigma$ contours for the emission and non–emission samples overlap only at small core radii.

### 5.3. Mass Estimates

Given the kinematic and spatial differences in the emission and non–emission galaxy samples, it is not surprising that the cluster masses they imply differ significantly. In particular, the projected mass estimator yields a total cluster mass of $M_{proj} = 1.5^{+0.4}_{-0.3} \times 10^{15} M_\odot$ for the 111 non–emission galaxies and $2.8^{+1.1}_{-0.7} \times 10^{15} M_\odot$ for the 58 emission galaxies (the 90% error ranges given assume that the mass uncertainty is dominated by the uncertainty in the velocity dispersion; see Heisler, Tremaine & Bahcall 1985). The entire sample of 221 galaxies within the velocity range of the cluster yields a mass of $2.9^{+0.5}_{-0.4} \times 10^{15} M_\odot$, consistent with the mass from the emission sample. The factor of 2 mass discrepancy between the emission and non–emission sample is roughly consistent with an infalling emission sample (see also Colless & Dunn 1995).

In Abell 576 we have carefully segregated the galaxy velocity sample into those with emission lines and those without. The typical procedure used to define cluster membership (e.g. den Hartog 1995) does not include this segregation, and this failure tends to increase optical velocity dispersion and virial mass estimates. Because emission lines are correlated with bluer color, the larger fraction of blue galaxy populations around distant clusters (Butcher & Oemler 1984) suggests that dispersions and virial masses of these clusters may be strongly affected by a failure to segregate the galaxy sample.

Figure 16 indicates that if an observer studying A576 measured 20 redshifts, the resulting cluster mass would also be sensitive to the magnitude range chosen for observation; the tendency to observe the brightest galaxies first would yield the lowest virial mass. The trend of increasing mass with magnitude is driven primarily by an $R$ band luminosity segregation. The brightest galaxies tend to be near the center of the cluster, while the faintest galaxies are distributed more homogeneously. It is probable that this apparent luminosity segregation is largely due to the morphology–density relation; the different kinematic, angular, and magnitude distributions of the emission and non–emission galaxies discussed in §3 strengthen this conclusion.

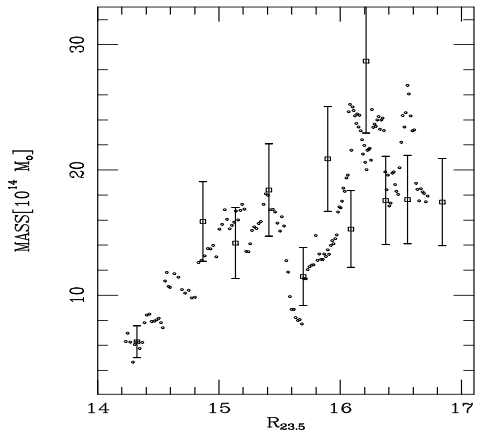

**Figure 16:** The variation in the cluster virial mass as a function of isophotal magnitude $R_{23.5}$. The kinematics of the fainter cluster population imply masses $\sim 3\times$ higher than the masses resulting from the brightest galaxies. The values with uncertainties are calculated using independent samples of 15 galaxies. The points without uncertainties represent values calculated using a sliding bin which enters from the bright end and progresses faintward. Masses of neighboring points without error bars are correlated. The first sliding bin point contains 15 galaxies, the second 16 up to a maximum of 30 galaxies per point. The final point on the faint end includes the 30 faintest galaxies.

Even after exclusion of the emission galaxies, it is difficult to calculate a consistent virial mass in A576. The projected mass implied by the 20 brightest members of the non–emission sample is less than the mass of the entire sample at 94% confidence ($M = 8.0^{+7.0}_{-3.0} \times 10^{14} M_\odot$); additional data will determine whether this $2\sigma$ variation is evidence for additional complication in the non–emission galaxy sample. Possible sources of mass variation with magnitude within the non–emission sample include contamination of the sample by non–virialized galaxies which exhibit no emission lines in their spectra or a simple variation of the orbital characteristics of the galaxy population with $R$ band magnitude (or morphological type).

Under the assumption that the X–ray emitting gas is in hydrostatic equilibrium, its distribution and temperature also yield a mass estimate (Bahcall & Sarazin 1977; Fabricant et al. 1980). In A576 we know the gas is not isothermal, but we do not have a measured temperature profile. Nevertheless, we estimate the cluster mass by considering the gas well outside the core where the temperature gradient is presumably small. Under these assumptions the enclosed cluster mass is determined primarily by the gas temperature $T$ and the radial falloff of the gas $\beta$:

$$M_X(<r) = \frac{kTr}{\mu m_p G} 3\beta \left(1 + \left(\frac{r_c}{r}\right)^2\right)^{-1} \simeq 2.8 \times 10^{14} M_\odot \left(\frac{T}{4 \text{ keV}} \frac{r}{\text{Mpc}}\right) \quad (5.3)$$

within a radius of 1 Mpc, a region comparable in size to that covered by the magnitude limited galaxy sample. In principle, the available X–ray spectral data provide only a lower limit on the temperature of the hot gas component ($> 4$ keV at 90% confidence). Thus, we write the mass in Equation 5.3 as a function of the gas temperature. The X–ray mass is less than the 90% lower limit of the optical virial mass (for the non–emission sample) unless the gas temperature is $\sim 17$ keV.

The mass of A576 appears to be poorly constrained even with our large dataset. How can we understand these mass discrepancies? For the optical data, the possible contamination of the non–emission sample by non–virialized cluster members or unresolved substructure could artifically boost the mass estimate. For example, excluding the 5 highest velocity galaxies which account for the skewness in the non–emission sample reduces the projected mass to $M = 9.3^{+2.6}_{-1.8} \times 10^{14} M_\odot$ and the velocity dispersion to $849^{+110}_{-85}$ km/s (90% confidence). In addition, the dynamical assumptions contained in the projected mass estimator may be invalid (e.g. The & White 1986). This possibility is supported by the variation among mass estimators; specifically, the ratio of the projected mass to the virial mass, median mass and mean mass (Heisler et al. 1985) for the entire sample of 111 non–emission velocities is 1:0.7:0.6:0.8. Excluding the 5 high velocity galaxies and applying the median mass estimator yields a "minimum" mass of $M_{min} = 7.0^{+1.9}_{-1.3} \times 10^{14} M_\odot$. This minimum virial mass equals the hydrostatic mass if the global gas temperature is $\sim 8$ keV. Because this intracluster medium temperature is consistent with the observational constraints (see §4.2), we consider this to be a likely resolution of the virial–hydrostatic mass discrepancy. Clearly, additional radial velocities, a spatially resolved ICM temperature, and a more sophisticated analysis of the optical mass estimates would yield an improved understanding of the nature of these mass discrepancies.

### 5.4. R Band Mass to Light Ratio

Using the measured mass range we constrain the cluster $R$ band mass–to–light ratio. With the absolute $R$ band magnitude of the sun ($M_R = 4.3$; Zombeck 1990) we calculate the total cluster luminosity using the best fit Schechter parameters for the background subtracted sample (§5.1). The result is $L_R = 1.3 \times 10^{12} h^{-2} L_\odot$, uncertain to $\sim 15\%$. Within the magnitude limited redshift sample, the non–emission galaxies contribute 75% of the cluster light. Because it is unlikely that the emission galaxies are part of the virialized population, we maintain that their light should not be included when calculating the cluster mass–to–light ratio. Thus, we calculate a corrected cluster luminosity of $L_R = 9.8 \times 10^{11} h^{-2} L_\odot$.

The mass uncertainty dominates the uncertainty in the mass–to–light ratio. The minimum virial mass presented above implies a cluster mass–to–light ratio of $760h$; given the range of hydrostatic and virial masses, the mass–to–light ratio could vary from $300h$ to $1600h$. These values are high, but include the range of values typical of groups and clusters (e.g. Ramella et al. 1989; David, Jones & Forman 1995). Recent weak lensing observations in distant clusters indicate mass–to–light ratios ranging from $\sim 300h$ (Tyson & Fischer 1995; Squires et al. 1996) to $\sim 700h$ (Fahlman et al. 1994; Carlberg, Yee & Ellingson 1994; Luppino & Kaiser 1996).

### 5.5. Baryon Fraction

A large measured gas mass fraction in clusters coupled with primordial nucleosynthesis calculations place strong constraints on the cosmological density parameter $\Omega_0$ (White *et al.* 1993; Steigman & Felten 1994; White & Fabian 1995). We calculate the gas mass fraction in A576 based on the central density and radial distribution of the gas along with estimates of the cluster virial mass. The mass in gas within $1h^{-1}$ Mpc is $9.4 \times 10^{12} h^{-2.5} M_\odot$; the value at $1.5h^{-1}$ Mpc is roughly a factor of 2 larger than the value in Jones & Forman (1984). We estimate a 20% uncertainty in the gas mass; contributions include uncertainties in the luminosity, the measured radial falloff parameters, and the unknown radial temperature behavior. Using the hydrostatic mass lower limit within 1 Mpc as the correct virial mass, we calculate a gas mass fraction

$$\left. \frac{M_{gas}}{M_X} \right|_{1h^{-1} \text{ Mpc}} = (0.032 \pm 0.006) h^{-\frac{3}{2}} \left( \frac{4 \text{ keV}}{T_X} \right) \quad (5.4)$$

The value at $0.5h^{-1}$ Mpc is $\sim 2.5 h^{-3/2}\%$. This gas baryon fraction is less than the fraction in 19 moderately distant clusters observed with the IPC; in this sample the gas baryon fraction varies from $3.5$–$7.8 h^{-3/2}\%$ (White & Fabian 1995). The hydrostatic mass we use here is a lower limit; the appropriate gas fraction for $T_X = 8$ keV is $\sim 1.6 h^{-3/2}\%$.

We explore the effect of changing the gas core radius on the baryon fraction because we are concerned that the low temperature gas in the core may lead to an underestimate of the core radius (Fabricant, Lecar & Gorenstein 1980). In particular, we calculate the baryon fraction with a core radius which is $240 h^{-1}$ kpc ($2\times$ larger than best fit value); the central density required to match the observed X–ray luminosity drops to $1.37 \times 10^{-3}$ cm$^{-3}$. The net effect of doubling the core radius is only a 10% increase in the gas mass.

Using the $R$ band luminosity we can estimate the baryon contribution from the galaxies. The mean global mass–to–light ratio in a sample of 41 elliptical galaxies studied by Lauer (1985) is $\langle M/L_B \rangle = 13.4h$ (with an RMS variation of $5.3h$). These mass–to–light ratios are transformations from the $R$ band; Lauer uses the average $B - R = 1.8$ color for the ellipticals in his sample for the transformation. We transform back using a $(B-R)_\odot = 1.18$; the resulting $\langle M/L_R \rangle = 7.6h$. We apply this mass–to–light ratio which is appropriate for ellipticals because we do not have morphological information in this cluster. We have corrected the cluster $R$ band luminosity for the contribution from the emission galaxies, but the non–emission sample contains lenticulars and some spirals in addition to the ellipticals. Because ellipticals are generally redder than both lenticulars and spirals, applying the elliptical mass–to–light ratio underestimates the total galaxy mass. Applying this average mass–to–light ratio to the corrected cluster $R$ band luminosity, we calculate a galaxy mass $M_{gal} \sim 7.4 \times 10^{12} h^{-1} M_\odot$, comparable to the gas mass within $1h^{-1}$ Mpc. This value is uncertain by $\sim 45\%$ because of (1) the 40% variation in the mass–to–light ratio over the population of ellipticals studied by Lauer (1985) and (2) the 15% uncertainty in the cluster $R$ band luminosity.

This galaxy mass implies a ratio of gas mass to galaxy mass of $M_{gas}/M_{gal} \sim 1.3h^{-3/2}$ (note that this is independent of the total cluster mass). This value is consistent with measurements in other clusters (David et al. 1990; Dell'Antonio, Geller, & Fabricant 1995). The total baryonic mass $M_b$ within $1h^{-1}$ Mpc yields a cluster baryon fraction

$$\left. \frac{M_b}{M_X} \right|_{1h^{-1} \text{ Mpc}} = \left( (0.032 \pm 0.006)h^{-\frac{3}{2}} + f(0.026 \pm 0.013) \right) \left( \frac{4 \text{ keV}}{T_X} \right) \quad (5.5)$$

where $f$ is the fraction of the galaxy mass which is baryonic. The ratio of total mass to baryonic mass ($M_{gas} + M_{gal}$) for $H_0 = 100$ km/s is $\sim 17(T_X/4 \text{ keV})$, about twice the value for four clusters of similar mass scale studied by David, Jones & Forman (1995).

In summary, although the relationship between the gas and galaxy mass in A576 is typical of other clusters, the mass fraction of these two components is smaller than in other clusters even when the cluster mass corresponding to a gas temperature of $T_X = 4$ keV (the 90% confidence lower limit) is assumed. If the minimum cluster virial mass is correct ($M_{min} = 7 \times 10^{14} M_\odot$, consistent with a gas temperature $T_X \sim 8$ keV) then the baryon fraction in A576 is $\sim 3\%$ (for $H_0 = 100$ km/s and galaxy baryon mass fraction $f = 1$; Equation 5.5).

## 6. CONCLUSIONS

With 281 redshifts, CCD galaxy photometry over a $2h^{-1}$ Mpc$\times 2h^{-1}$ Mpc region centered on the cluster, and the available X–ray data, we study the cluster dynamics and galaxy population in A576. We focus on an 86% complete magnitude limited sample ($R_{23.5} < 17$) of 204 galaxies with measured redshifts; 169 of these galaxies have velocities within the cluster range.

Within the magnitude limited sample of cluster galaxies, 58 galaxies have emission lines in their spectra and 111 do not; the spatial, kinematic and magnitude distributions of the galaxies with and without emission lines are significantly different (see §3). Similar kinematic differences have been noted as a function of galaxy morphology in the Virgo cluster (Huchra 1985) and as a function of galaxy color in the Coma cluster (Colless & Dunn 1995).

The rise in the velocity dispersion of the emission line population toward the cluster core (98% confidence) indicates that this population is dynamically "aware" of the cluster and probably falling into the cluster for the first time. The observations in A576 favor a model where the cluster environment transforms the gas rich population surrounding the cluster into the gas poor population clustered in the core.

The data in A576 indicate (§4) that the cluster core contains a low velocity dispersion population of non–emission galaxies and the previously known (White & Silk 1980; Rothenflug et al. 1984) low temperature gas ($\sigma_{core} = 387^{+250}_{-105}$ km/s and $T_X = 1.6^{+0.4}_{-0.3}$ keV at 90% confidence). These values are significantly lower than the global non–emission velocity dispersion of $\sigma = 977^{+124}_{-96}$ km/s and the global intracluster medium temperature of $T_X > 4$ keV. We present two heuristic models; in one model relaxation and cooling effects which are more efficient in the core lead to the emergence of cool core populations, and in the second model the core of A576 contains the remnants of a low mass subcluster. Because the galaxies in the low dispersion core span a range in magnitude similar to the global sample, and because there is only weak evidence for a cooling flow in the intracluster medium ($t_{cool} \sim 2 \times 10^{10}$ yrs at a radius of $55h^{-1}$ kpc, $1 \times 10^{10}$ yrs in the center and no X–ray emission excess), we argue that the subcluster remnant model is favored. We suggest additional observations to discriminate clearly between these two models. If the subcluster remnant model is valid, substructure relaxation timescales in the cores of clusters require closer examination.

We fit a Schechter function (§5.1) to the galaxy magnitude distribution to $M_R = -16.5$. The best fit parameters (and 90% confidence intervals) are $M_* = -20.8^{+0.6}_{-0.3}$ and $\alpha = -1.18^{+0.13}_{-0.11}$, but the quality of the fit is poor (reduced $\chi^2 = 3$ for 22 degrees of freedom). We study the radial galaxy distribution and demonstrate that with ~100 known cluster galaxies, the best fit core radius and radial fall–off are not tightly constrained (§5.2).

The cluster mass in A576 is poorly constrained (§5.3). The lower limit of the hydrostatic mass $M_X = 2.8 \times 10^{14} M_\odot (T_X/4 \text{ keV})$ within $1h^{-1}$ Mpc is far lower than the projected mass estimate $M_V = 1.5^{+0.4}_{-0.3} \times 10^{15} M_\odot$ obtained using the non–emission sample of galaxies. The mass discrepancy may be resolved through a combination of a high intracluster medium temperature ($T_X \sim 8$ keV– consistent with observational constraints) and unresolved substructure or contamination from individual non–virialized galaxies. In addition, we demonstrate that in A576 the failure to exclude the unclustered emission population results in an 18% increase in the cluster dispersion and a factor of two increase in the cluster virial mass. Assuming a correlation between a bluer color and the presence of emission lines in galaxy spectra, we note that the Butcher–Oemler effect (Butcher & Oemler 1984) implies that "contamination" by gas rich galaxies may more significantly affect estimates of cluster dispersions and virial masses in high redshift clusters.

Finally, we use the hydrostatic mass as a lower limit to investigate the cluster baryon fraction (§5.5). The gas mass fraction ($\sim 0.03h^{-3/2}(4 \text{ keV}/T_X)$) is lower than that found ($0.035$–$0.078h^{-3/2}$) in a sample of 19 clusters observed with the *Einstein* IPC (White & Fabian 1995). If the majority of the galaxy mass is baryonic then the total baryon fraction within the cluster is $\sim 6(4 \text{ keV}/T_X)\%$. If the true cluster mass is closer to the virial mass than to the hydrostatic mass (lower limit), then the cluster baryon crisis (White *et al.* 1993; Steigman & Felten 1994) disappears in A576.

## ACKNOWLEDGEMENTS

We wish to thank Mary Crone for helping with one Decaspec run, Larry David for useful discussions and for providing the reduced MPC spectrum of A576, and John Huchra for taking a few snapshots of A576 to replace some high extinction images in our mosaic. Thanks go to Gus Evrard, Marijn Franx, and an anonymous referee for helpful comments. JJM was supported by a NASA graduate student research fellowship. This research was supported in part by NAGW–201.

## REFERENCES

Abell, G.O. 1958, ApJS, 3, 211
Bahcall, N.A. 1975, ApJ, 198, 249
Bahcall, J.N. & Sarazin, C.L. 1977, ApJ, 213, L99
Beers, T.C., Forman, W., Huchra, J.P., Jones, C. & Gebhardt, K. 1991, AJ, 102, 1581
Binggeli, B, Tammann, G.A. & Sandage, A. 1987, AJ, 94, 251
Binggeli, B., Sandage, A. & Tammann, G.A. 1988, *Annual Reviews of Astronomy and Astrophysics*, 26, 509
Binney, J. 1977, MNRAS, 181, 735
Binney, J. & Tremaine, S. 1987, *Galactic Dynamics*, Princeton University Press: Princeton
Bothun, G.D. & Dressler, A. 1986, ApJ, 301, 57
Bothun, G.D. & Schombert, J.M 1990, ApJ, 360, 436
Briel, U.G. & Henry, J.P. 1995, Nature, 372, 439
Burns, J.O., Roettiger, K., Ledlow, M. & Klypin, A. 1994, ApJ, 427, L87
Burns, J.O., Roettiger, K., Pinkney, J., Perley, R.A., Owen, F.N. & Voges, W. 1995, ApJ, 446, 583
Burstein, D. & Heiles, C. 1978, ApJ, 225, 40
Butcher, H. & Oemler, A.J. 1978, ApJ, 226, 559
Butcher, H. & Oemler, A. 1984, ApJ, 285, 426
Capelato, H.G., Gerbal, D., Mathez, G., Mazure, A., Salvador-Sole, E. & Sol, H. 1980, ApJ, 241, 521
Carlberg, R.G., Yee, H.K.C. & Ellingson, E. 1994, ApJ, 437, 63
Chandrasekhar, S. 1942, *Principles of Stellar Dynamics*, (University of Chicago Press: Chicago)
Colless, M. & Dunn, A. 1995, ApJ, in press and SISSA preprint 9508070
Couch, W.J., Ellis, R.S., Sharples, R.M. & Smail, I. 1994, ApJ, 430, 121
Cowie, L.L & McKee, C.F. 1977, ApJ, 211, 135
Cowie, L.L. & Songaila, A. 1977, Nature, 266, 501
Cowie, L. L. & Hu, E.M. 1986, ApJ, 305, L39
Crone, M.M. & Geller, M.J. 1995, AJ, 110, 21
Danese, L. De Zotti, G. & di Tullio, G. 1980, AA, 82, 322
David, L.P. 1995, private communication
David, L.P., Arnaud, K.A., Forman, W. & Jones, C. 1990, ApJ, 356, 32
David, L.P., Jones, C. & Forman, W. 1995, ApJ, 445, 578
David, L.P., Arnaud, K.A., Forman, W. & Jones, C. 1990, ApJ, 356, 32
David, L.P., Slyz, A., Jones, C., Forman, W., Vrtilek, S.D. & Arnaud, K.A. 1993, ApJ, 412, 479
Dell'Antonio, I.P., Geller, M.J. & Fabricant, D.G. 1995, AJ, 110, 502
den Hartog, R. 1995, PhD thesis, Leiden Observatory
Dressler, A. 1978, ApJ, 223, 765
Dressler, A. 1978, ApJ, 226, 55
Dressler, A. 1980, ApJ, 236, 351
Dressler, A. & Gunn, J.E. 1982, ApJ, 263, 533
Dressler, A. & Gunn, J.E. 1983, ApJ, 270, 7
Dressler, A., Oemler, A., Butcher, H.R. & Gunn, J.E. 1994, ApJ, 430, 107
Dressler, A. & Shectman, S. 1988, AJ, 95, 985
Edge, A.C., Stewart, G.C. & Fabian, A.C. 1992, MNRAS, 258, 177
Evrard, A.E. 1990, ApJ, 363, 349
Evrard, A.E. 1991, MNRAS, 248, 8
Evrard, A.E., Mohr, J.J., Fabricant, D.G. & Geller, M.J. 1993, ApJ, 419, L9
Fabricant, D.G., Beers, T.C., Geller, M.J., Gorenstein, P., Huchra, J.P. & Kurtz, M.J. 1986, ApJ, 308, 530
Fabricant, D.G. & Hertz, E. 1990, SPIE Proc. 1235, 747
Fabricant, D.G., Kent, S.M. & Kurtz, M.J. 1989, ApJ, 336, 77
Fabricant, D.G., Kurtz, M., Geller, M.J., Zabludoff, A.I., Mack, P. & Wegner, G. 1993, AJ, 105, 788
Fabricant, D.G., Lecar, M. & Gorenstein, P. 1980, ApJ, 241, 552
Fahlman, G., Kaiser, N., Squires, G. & Woods, D. 1994, ApJ, 437, 56
Forman, W. Bechtold, J., Blair, W., Giacconi, R., van Speybroeck, L. & Jones, C. 1981, ApJ, 243, 133
Frenk, C.S., Evrard, A.E., White, S.D.M. & Summers, F.J. 1995, preprint.
Geller, M.J. *et al.* 1995, in preparation

Geller, M.J. & Beers, T.C. 1982, PASP, 94, 421
Grindlay, J.E., Marshall, H., Hertz, P. Soltan, A., Weisskopf, M., Elsner, R., Ghosh, P., Darbro, W., & Sutherland, P.G. 1980, ApJ, 240, L121
Gunn, J.E. & Gott, J.R. 1972, ApJ, 176, 1
Heisler, J., Tremaine, S. & Bahcall, J.N. 1985, ApJ, 298, 8
Henry, J.P. & Briel, 1995, ApJ, 443, 9
Hill, J.M., Angel, R.P., Scott, J.S., Lindley, D. & Hintzen, P. 1980, ApJ, 242, L69
Hintzen, P., Hill, J.M., Lindley, D., Scott, J.S. & Angel, J.R.P. 1982, AJ, 87, 1656
Huchra, J. P., in proceedings of the *ESO Workshop on the Virgo Cluster of Galaxies*, Garching: European Southern Observatory, 1985
Huchra, J., Geller, M.J., CLemens, C., Tokarz, S. & Michel, A. 1992, Bull CDS, 41, 31
Jarvis, J. & Tyson, J. 1981, AJ, 86, 476
Jones, C. & Forman, W. 1984, ApJ, 276, 38
Jones, C. & Forman, W. 1992, in Clusters and Superclusters of Galaxies (NATO ASI Vol. 366), ed. A.C. Fabian, London: Kluwer
Jones, L.R., Fong, R., Shanks, T., Ellis, R.S. & Peterson, B.A. 1991, MNRAS, 249, 481
Jørgensen, I., Franx, M. & Kjærgaard, P. 1993, ApJ, 411, 34
Kent, S.M. & Gunn, J.E. 1982, AJ, 87, 945
Kurtz, M. *et al.* 1995, private communication
Lacey, C. & Cole, S. 1993, MNRAS, 262, 627
Landolt, A. 1992, AJ, 104, 340
Lauer, T.R. 1985, ApJ, 292, 104
Lea, S.M., Silk, J., Kellogg, E. & Murray, S. 1973, ApJ, 184, L105
Lin, H. Kirshner, R.P., Shectman, S.A., Landy, S.D., Oemler, A., Tucker, D.L. & Schechter, P.L. 1995, preprint
Loeb, A. & Mao, S. 1994, ApJ, 435, L109
Lopez–Cruz, O. 1995, PhD thesis, University of Toronto
Lubin, L.M. & Bahcall, N.A. 1993, ApJ, 415, L17
Lugger, P.M. 1985, ApJ, 303, 535
Luppino, G.A. & Kaiser, N. 1996, preprint
Marzke, R.O., Huchra, J.P. & Geller, M.J. 1994, ApJ, 428, 43
Marzke, R., Huchra, J. & Geller, M.J. 1995, AJ, submitted
McLeod, B. 1995, private communication
Melnick, J. & Sargent, W. 1977, ApJ, 215, 401
Merrifield, M.R. & Kent, S.M. 1991, AJ, 101, 783
Merritt, D. 1984, ApJ, 276, 26
Merritt, D. 1985, ApJ, 289, 18
Metcalfe, N., Shanks, T., Fong, R. & Jones, L.R. 1991, MNRAS, 249, 498
Miralda–Escudé, J. & Babul, A. 1994, ApJ, submitted
Mohr, J.J., Fabricant, D.G. & Geller, M.J. 1993, ApJ, 413, 492
Mohr, J.J., Evrard, A.E., Fabricant, D.G. & Geller, M.J. 1995, ApJ, 447, 8
Oegerle, W.R., Hoessel, J.G. & Jewison, M.S. 1987, AJ, 93, 519
Oegerle, W.R. & Hill, J.M. 1994, AJ, 107, 857
Oemler, A. 1992, in *Clusters and Superclusters of Galaxies*, ed. A.C. Fabian, (Dordrecht: Kluwer), 29
Pinkney, J., Rhee, G., Burns, J.O., Hill, J.M., Oegerle, W., Batuski, D. & Hintzen, P. 1993, ApJ, 416, 36
Postman, M. & Geller, M.J. 1984, ApJ, 281, 95
Quintana, H. 1979, AJ, 84, 15
Ramella, M., Geller, M.J., & Huchra, J.P. 1989, ApJ, 344, 57
Regös, E. & Geller, M.J. 1989, AJ, 98, 755
Richstone, D.O., Loeb, A. & Turner, E.L. 1992, ApJ, 393, 477
Rieke, G.H & Lebofsky, M.J. 1985, ApJ, 288, 618
Rothenflug, R., Vigrouw, L., Mushotsky, R.F. & Holt, S.S. 1984, ApJ, 279, 53
Sarazin, C.L. 1988, *X-ray Emissions from Clusters of Galaxies*, Cambridge University Press: Cambridge

Schechter, P. 1976, ApJ, 203, 297

Sodré, L., Capelato, H.V., Steiner, J.E. & Mazure, A. 1989, AJ, 97, 1279

Sodré, L., Capelato, H.V., Steiner, J.E., Proust, D. Mazure, A. 1992, MNRAS, 259, 233

Soltan, A, & Fabricant, D.G. 1990, ApJ, 364, 433

Spitzer, L. Jr. 1956, *Physics of Fully Ionized Gases*, (Interscience: New York)

Spitzer, L. & Hart, M.H. 1971, ApJ, 164, 399

Squires, G., Kaiser, N., Fahlman, G., Babul, A. & Woods, D. 1996, preprint

Steigman, G. & Felten, J.E. 1994, preprint

The, L.S. & White, S.D.M. 1986, AJ, 92, 1248

Thorstensen, J. 1993, private communication

Tyson, J.A. & Fischer, P. 1995, ApJ, 446, L55

Valdes, F. 1982, SPIE Proc., 331, 465

van Haarlem, M. 1992, PhD thesis, Leiden Observatory

Van Speybroeck, L., Szczypek, A. & Fabricant, D.G. 1980, SAO Special Report 393, Harnden, F.R., Fabricant, D.G. Harris, D.E. & Schwarz, J. pg 234

Visvanathan, N. & Sandage, A. 1977, ApJ, 216, 214

White, D.A. & Fabian, A.C. 1995, MNRAS, 273, 72

White, S.D.M. 1978, MNRAS, 183, 341

White, S.D.M. & Silk, J. 1980, ApJ, 241, 864

White, S.D.M., Navarro, J.F., Evrard, A.E. & Frenk, C.S. 1993, Nature, 366, 429

Zabludoff, A.I. & Franx, M. 1993, AJ, 106, 1314

Zabludoff, A.I. & Zaritsky, D. 1995, ApJ, 447, L21

Zombeck, M.V. 1990, *Handbook of Space Astronomy and Astrophysics*, Cambridge University Press: Cambridge

Appendix: Galaxy Velocities and Photometry

| RA (1950) | Decl | $R_{23.5}$ | $\sigma_R$ | $v$ | $\sigma_v$ | T | RA (1950) | Decl | $R_{23.5}$ | $\sigma_R$ | $v$ | $\sigma_v$ | T |
|---|---|---|---|---|---|---|---|---|---|---|---|---|---|
| 7 18 06.52 | 55 51 49.8 | 13.95 | 0.021 | 11395† | 100 | N | 7 19 10.95 | 55 21 03.2 | 16.04 | 0.033 | 22536 | 49 | N |
| 7 17 39.22 | 55 29 01.4 | 14.00 | 0.033 | 10667 | 34 | N | 7 20 53.94 | 55 39 28.4 | 16.04 | 0.035 | 11547 | 50 | E |
| 7 19 02.67 | 55 43 25.3 | 14.09 | 0.023 | 12145† | 100 | N | 7 21 05.22 | 56 19 27.0 | 16.04‡ | 0.250 | 12524 | 65 | E |
| 7 16 07.76 | 55 36 50.0 | 14.22 | 0.020 | 12100 | 51 | N | 7 12 15.07 | 56 03 46.5 | 16.05‡ | 0.250 | 11526 | 61 | N |
| 7 16 30.66 | 55 51 53.9 | 14.22 | 0.029 | 12524† | 100 | N | 7 15 00.97 | 55 16 53.3 | 16.05‡ | 0.250 | 38736 | 38 | N |
| 7 17 38.47 | 55 46 27.1 | 14.25 | 0.037 | 9847† | 100 | N | 7 20 11.60 | 55 47 49.8 | 16.06 | 0.033 | 10627 | 48 | E |
| 7 19 01.93 | 55 07 54.8 | 14.27‡ | 0.250 | 11158 | 40 | E | 7 20 26.54 | 55 26 35.8 | 16.06 | 0.037 | 11343 | 52 | E |
| 7 18 00.37 | 55 58 13.5 | 14.28 | 0.080 | 11993† | 100 | N | 7 16 43.65 | 55 49 06.4 | 16.07 | 0.031 | 12598 | 25 | N |
| 7 17 24.06 | 55 51 22.6 | 14.29 | 0.021 | 11436† | 100 | N | 7 12 32.60 | 55 26 55.9 | 16.08‡ | 0.250 | 4695 | 70 | E |
| 7 16 18.06 | 56 29 01.7 | 14.30‡ | 0.250 | 13547 | 38 | N | 7 17 35.61 | 55 55 53.0 | 16.08 | 0.021 | 9834† | 100 | N |
| 7 15 56.84 | 56 01 28.9 | 14.33 | 0.024 | 12242 | 42 | E | 7 17 21.15 | 55 33 34.6 | 16.09 | 0.022 | 10806 | 34 | N |
| 7 17 15.13 | 55 54 18.9 | 14.37 | 0.029 | 11016† | 100 | N | 7 18 00.21 | 56 15 05.6 | 16.09 | 0.017 | 10074 | 40 | E |
| 7 14 46.01 | 55 33 33.0 | 14.39 | 0.080 | 10662 | 45 | N | 7 16 42.16 | 55 45 19.6 | 16.10 | 0.031 | 11316 | 35 | N |
| 7 17 30.04 | 55 40 55.9 | 14.46 | 0.037 | 11501† | 100 | N | 7 15 14.72 | 55 46 46.6 | 16.11 | 0.020 | 9805 | 42 | E |
| 7 17 15.59 | 55 53 34.2 | 14.50 | 0.029 | 12103† | 100 | N | 7 15 49.88 | 55 47 37.3 | 16.11 | 0.019 | 13232 | 43 | N |
| 7 15 22.55 | 55 42 04.0 | 14.51 | 0.019 | 9912 | 47 | N | 7 20 54.28 | 56 11 43.5 | 16.11 | 0.035 | 12280 | 35 | E |
| 7 12 48.10 | 55 17 32.5 | 14.53‡ | 0.250 | 11491 | 38 | E | 7 20 46.13 | 55 22 29.0 | 16.12 | 0.037 | 25991 | 33 | N |
| 7 15 04.22 | 56 09 32.7 | 14.56 | 0.024 | 11971 | 27 | E | 7 17 44.10 | 56 13 29.7 | 16.13 | 0.016 | 11605 | 37 | E |
| 7 14 42.95 | 55 34 35.5 | 14.62 | 0.038 | 11633* | 39 | E | 7 16 13.34 | 56 06 08.6 | 16.14 | 0.027 | 12630 | 34 | N |
| 7 15 48.70 | 55 39 54.4 | 14.68 | 0.030 | 12696 | 43 | N | 7 16 55.58 | 55 50 19.9 | 16.14 | 0.031 | 9630 | 69 | E |
| 7 13 57.45 | 55 09 05.6 | 14.69‡ | 0.250 | 12850 | 40 | N | 7 17 08.76 | 55 49 33.1 | 16.14 | 0.031 | 11127 | 57 | N |
| 7 16 16.89 | 55 55 35.0 | 14.69 | 0.030 | 9536 | 87 | E | 7 17 19.24 | 56 09 46.5 | 16.17 | 0.030 | 10704 | 33 | N |
| 7 19 18.11 | 55 38 05.3 | 14.73 | 0.030 | 10218† | 100 | N | 7 15 43.02 | 55 19 21.1 | 16.18 | 0.043 | 11636 | 38 | N |
| 7 20 28.32 | 55 41 52.0 | 14.78 | 0.033 | 11665† | 100 | N | 7 16 38.58 | 56 03 32.3 | 16.18 | 0.027 | 14854 | 41 | E |
| 7 16 52.07 | 55 49 20.0 | 14.81 | 0.031 | 11020† | 100 | N | 7 18 04.22 | 56 18 19.4 | 16.18 | 0.018 | 10140 | 41 | E |
| 7 17 34.47 | 55 47 07.2 | 14.85 | 0.037 | 19857† | 100 | N | 7 17 37.34 | 56 09 26.1 | 16.19 | 0.031 | 13018 | 53 | E |
| 7 17 58.95 | 55 45 52.2 | 14.86 | 0.037 | 10268† | 100 | N | 7 19 38.31 | 56 04 58.5 | 16.19 | 0.023 | 19787 | 40 | N |
| 7 15 37.20 | 55 30 01.2 | 14.87 | 0.030 | 10542 | 47 | N | 7 15 00.05 | 55 35 57.7 | 16.20 | 0.031 | 14361 | 42 | E |
| 7 16 32.00 | 56 04 16.1 | 14.88 | 0.027 | 12153× | 100 | N | 7 15 48.30 | 55 52 54.0 | 16.21 | 0.032 | 11072 | 49 | N |
| 7 15 44.44 | 55 42 09.0 | 14.91 | 0.019 | 11792 | 30 | E | 7 18 28.77 | 55 45 43.8 | 16.21 | 0.038 | 11985 | 39 | N |
| 7 12 24.93 | 55 28 39.1 | 14.92‡ | 0.250 | 14517 | 41 | E | 7 15 36.17 | 56 04 05.6 | 16.22 | 0.024 | 12481 | 40 | E |
| 7 16 57.10 | 55 45 30.6 | 14.93 | 0.031 | 11224† | 100 | N | 7 15 41.07 | 55 35 00.0 | 16.22 | 0.031 | 10165 | 38 | E |
| 7 20 25.16 | 55 20 08.5 | 14.94 | 0.036 | 14452 | 47 | N | 7 16 20.47 | 56 21 58.3 | 16.22‡ | 0.250 | 30222 | 31 | N |
| 7 15 32.49 | 55 31 39.3 | 14.95 | 0.030 | 19671 | 49 | E | 7 20 13.58 | 56 21 50.1 | 16.22 | 0.035 | 13597 | 70 | E |
| 7 17 44.52 | 56 18 30.7 | 14.95 | 0.016 | 10127 | 40 | E | 7 16 59.52 | 55 59 59.3 | 16.24 | 0.030 | 12790† | 100 | N |
| 7 16 47.64 | 55 48 56.8 | 14.97 | 0.031 | 11385† | 100 | N | 7 16 13.51 | 55 55 28.6 | 16.25 | 0.030 | 11445 | 25 | N |
| 7 17 38.00 | 56 17 48.5 | 14.97 | 0.016 | 13096† | 100 | N | 7 12 05.68 | 55 18 42.1 | 16.28‡ | 0.250 | 11124 | 47 | E |
| 7 18 49.06 | 55 44 38.0 | 14.97 | 0.023 | 12276† | 100 | N | 7 21 14.45 | 55 36 29.9 | 16.28‡ | 0.250 | 10639 | 36 | N |
| 7 12 35.60 | 55 33 35.2 | 14.99‡ | 0.250 | 11254 | 44 | E | 7 15 58.59 | 55 54 01.1 | 16.29 | 0.032 | 11009 | 44 | N |
| 7 18 45.41 | 55 38 08.4 | 15.00 | 0.030 | 10899† | 100 | N | 7 16 19.18 | 55 45 38.0 | 16.30 | 0.031 | 11113 | 35 | N |
| 7 16 30.96 | 55 44 37.3 | 15.01 | 0.031 | 11285 | 40 | N | 7 19 52.13 | 55 20 29.3 | 16.30 | 0.037 | 25860 | 35 | N |
| 7 17 26.39 | 55 51 07.4 | 15.01 | 0.021 | 12177† | 100 | N | 7 19 47.20 | 56 21 18.3 | 16.32 | 0.035 | 12906 | 36 | E |
| 7 15 26.48 | 56 19 19.3 | 15.03 | 0.040 | 10910 | 47 | E | 7 17 33.72 | 55 52 52.2 | 16.33 | 0.021 | 11150 | 54 | N |
| 7 16 13.08 | 55 58 41.1 | 15.06 | 0.029 | 11891 | 36 | N | 7 16 40.77 | 55 43 05.0 | 16.34 | 0.031 | 13087 | 52 | E |
| 7 19 26.77 | 55 32 15.4 | 15.08 | 0.030 | 11325 | 60 | E | 7 14 16.21 | 55 32 30.1 | 16.35 | 0.039 | 13094 | 50 | E |
| 7 20 10.94 | 55 41 15.8 | 15.10 | 0.033 | 13605 | 43 | N | 7 19 30.51 | 55 18 58.3 | 16.35 | 0.033 | 11053 | 31 | N |
| 7 22 17.50 | 56 23 24.9 | 15.10‡ | 0.250 | 13442 | 43 | N | 7 19 54.39 | 55 22 21.4 | 16.35 | 0.037 | 12462 | 43 | N |
| 7 19 13.46 | 55 24 18.7 | 15.12 | 0.033 | 11975 | 37 | N | 7 17 46.23 | 55 54 40.8 | 16.36 | 0.022 | 10680† | 100 | N |
| 7 18 03.99 | 55 45 33.9 | 15.15 | 0.037 | 12494† | 100 | N | 7 15 05.32 | 55 37 08.1 | 16.37 | 0.031 | 19508 | 45 | E |

Velocities and Photometry continued

| RA (1950) | Decl | $R_{23.5}$ | $\sigma_R$ | $v$ | $\sigma_v$ | T | RA (1950) | Decl | $R_{23.5}$ | $\sigma_R$ | $v$ | $\sigma_v$ | T |
|---|---|---|---|---|---|---|---|---|---|---|---|---|---|
| 7 17 11.91 | 56 01 02.7 | 15.16 | 0.027 | 13021† | 100 | N | 7 18 34.32 | 55 50 13.7 | 16.38 | 0.024 | 14742 | 47 | E |
| 7 20 22.41 | 55 39 59.9 | 15.16 | 0.033 | 11619† | 100 | N | 7 21 51.58 | 56 12 29.5 | 16.38‡ | 0.250 | 11059 | 42 | E |
| 7 14 23.75 | 55 28 40.1 | 15.18 | 0.036 | 18006 | 35 | N | 7 15 48.22 | 56 33 41.6 | 16.40‡ | 0.250 | 14999 | 69 | N |
| 7 16 31.35 | 55 22 58.6 | 15.19 | 0.021 | 19840 | 43 | E | 7 15 43.35 | 56 10 33.2 | 16.42 | 0.024 | 10947 | 42 | E |
| 7 20 00.07 | 56 30 41.8 | 15.19‡ | 0.250 | 13306 | 39 | E | 7 19 17.37 | 55 27 59.5 | 16.43 | 0.034 | 12027 | 34 | N |
| 7 20 54.28 | 55 58 36.9 | 15.20 | 0.039 | 18047† | 100 | N | 7 20 38.84 | 56 20 02.0 | 16.43 | 0.035 | 11693 | 57 | E |
| 7 17 50.18 | 55 56 06.6 | 15.22 | 0.021 | 11080† | 100 | N | 7 15 31.39 | 55 31 27.2 | 16.46 | 0.031 | 19855 | 38 | N |
| 7 16 51.96 | 55 26 41.4 | 15.24 | 0.021 | 20360 | 50 | E | 7 17 08.52 | 55 54 46.0 | 16.46 | 0.030 | 10254 | 27 | N |
| 7 17 29.26 | 56 15 59.8 | 15.25 | 0.016 | 10943† | 100 | N | 7 18 24.18 | 56 16 19.2 | 16.48 | 0.018 | 11496 | 58 | E |
| 7 21 43.60 | 55 35 38.8 | 15.26‡ | 0.250 | 10693 | 31 | N | 7 13 14.89 | 55 21 37.9 | 16.49‡ | 0.250 | 10750 | 50 | E |
| 7 14 10.65 | 55 49 25.4 | 15.27 | 0.028 | 14066 | 39 | N | 7 13 48.44 | 56 00 11.1 | 16.49 | 0.057 | 12494 | 40 | E |
| 7 17 22.97 | 56 31 58.8 | 15.28‡ | 0.250 | 10274 | 36 | N | 7 16 09.13 | 56 24 59.8 | 16.49‡ | 0.250 | 11163 | 97 | N |
| 7 16 15.15 | 56 28 42.1 | 15.30‡ | 0.250 | 14147 | 44 | N | 7 18 05.48 | 55 48 46.3 | 16.49 | 0.038 | 13337 | 51 | N |
| 7 17 29.01 | 55 59 09.5 | 15.30 | 0.021 | 13268† | 100 | N | 7 19 19.21 | 55 55 10.5 | 16.49 | 0.040 | 11187 | 41 | N |
| 7 19 04.67 | 56 04 22.1 | 15.31 | 0.022 | 13550 | 55 | E | 7 19 57.62 | 56 19 36.6 | 16.50 | 0.035 | 11932 | 45 | N |
| 7 21 41.40 | 56 04 36.9 | 15.31‡ | 0.250 | 10105 | 38 | E | 7 18 56.41 | 55 29 03.9 | 16.51 | 0.034 | 12117 | 69 | E |
| 7 12 43.55 | 56 02 17.8 | 15.32‡ | 0.250 | 13360 | 54 | E | 7 12 01.51 | 55 26 56.9 | 16.53‡ | 0.250 | 14368 | 45 | E |
| 7 20 32.54 | 56 12 11.6 | 15.32 | 0.034 | 10559 | 28 | N | 7 14 42.01 | 56 11 08.7 | 16.53 | 0.032 | 11477 | 46 | E |
| 7 17 23.94 | 55 48 21.7 | 15.33 | 0.037 | 20216† | 100 | N | 7 17 50.48 | 55 49 07.6 | 16.53 | 0.038 | 9713 | 55 | E |
| 7 18 35.43 | 55 55 37.3 | 15.33 | 0.039 | 10990† | 100 | N | 7 18 14.68 | 55 40 43.2 | 16.53 | 0.038 | 13052 | 43 | E |
| 7 15 26.28 | 56 18 56.5 | 15.35 | 0.040 | 10666 | 58 | N | 7 17 16.94 | 55 45 44.1 | 16.54 | 0.031 | 12641 | 40 | N |
| 7 15 46.69 | 55 49 59.6 | 15.35 | 0.019 | 12465 | 29 | N | 7 19 10.86 | 56 32 58.9 | 16.55‡ | 0.250 | 11483 | 46 | N |
| 7 12 54.05 | 55 18 00.9 | 15.36‡ | 0.250 | 11502 | 38 | N | 7 17 36.73 | 56 01 24.9 | 16.56 | 0.031 | 9592 | 49 | E |
| 7 15 13.54 | 55 50 05.3 | 15.36 | 0.019 | 19836 | 38 | N | 7 19 54.78 | 56 00 18.9 | 16.56 | 0.040 | 11306 | 56 | E |
| 7 17 38.95 | 55 46 05.6 | 15.37 | 0.037† | 10011 | 100 | N | 7 12 39.53 | 55 37 54.9 | 16.57‡ | 0.250 | 26130 | 49 | E |
| 7 20 46.16 | 55 50 48.4 | 15.38 | 0.039 | 12427 | 43 | E | 7 18 14.39 | 55 46 23.6 | 16.57 | 0.038 | 10204 | 33 | N |
| 7 17 59.92 | 55 25 18.9 | 15.39 | 0.033 | 11082 | 36 | N | 7 16 30.70 | 56 24 23.3 | 16.59‡ | 0.250 | 14381 | 38 | N |
| 7 19 25.28 | 56 21 35.9 | 15.39 | 0.027 | 12861† | 100 | N | 7 21 08.55 | 56 15 23.8 | 16.59‡ | 0.250 | 18222 | 42 | E |
| 7 21 40.03 | 55 42 55.7 | 15.41‡ | 0.250 | 10740 | 58 | E | 7 14 39.28 | 55 26 58.7 | 16.60 | 0.037 | 10714 | 67 | E |
| 7 17 25.72 | 55 54 29.3 | 15.42 | 0.021 | 11867† | 100 | N | 7 21 41.39 | 55 32 44.7 | 16.60‡ | 0.250 | 11377 | 59 | E |
| 7 15 29.61 | 55 22 42.8 | 15.44 | 0.043 | 10756 | 52 | N | 7 18 37.62 | 55 41 19.5 | 16.61 | 0.024 | 22362 | 76 | N |
| 7 16 42.59 | 55 52 13.0 | 15.44 | 0.030 | 11415† | 100 | N | 7 18 59.60 | 55 41 13.2 | 16.64 | 0.024 | 12103 | 57 | N |
| 7 19 26.74 | 56 05 26.0 | 15.45 | 0.023 | 10903 | 61 | N | 7 14 39.26 | 55 43 24.9 | 16.65 | 0.029 | 9875 | 49 | E |
| 7 17 08.51 | 55 59 02.5 | 15.46 | 0.030 | 9845† | 100 | N | 7 15 30.25 | 56 19 58.8 | 16.66 | 0.041 | 13620 | 41 | N |
| 7 18 00.64 | 56 02 19.7 | 15.47 | 0.030 | 9930 | 25 | N | 7 15 45.20 | 56 05 00.7 | 16.69 | 0.025 | 40579 | 25 | N |
| 7 13 17.50 | 55 28 49.6 | 15.48‡ | 0.250 | 14255 | 33 | N | 7 17 00.58 | 55 59 17.8 | 16.69 | 0.030 | 9462† | 100 | N |
| 7 18 54.02 | 55 41 59.4 | 15.49 | 0.023 | 22641 | 47 | N | 7 15 55.07 | 56 06 31.9 | 16.70 | 0.024 | 12113 | 36 | N |
| 7 20 27.46 | 56 30 08.1 | 15.49‡ | 0.250 | 11025 | 55 | E | 7 16 18.23 | 55 42 52.6 | 16.70 | 0.031 | 12216 | 62 | E |
| 7 21 14.69 | 55 36 41.2 | 15.49‡ | 0.250 | 10564† | 100 | N | 7 20 41.74 | 56 00 33.6 | 16.72 | 0.040 | 17967 | 43 | N |
| 7 21 06.95 | 55 20 08.2 | 15.52‡ | 0.250 | 11467 | 28 | N | 7 17 02.91 | 55 41 53.9 | 16.73 | 0.032 | 37858 | 45 | N |
| 7 18 21.63 | 55 32 48.9 | 15.54 | 0.022 | 11301 | 44 | N | 7 16 33.38 | 56 17 53.6 | 16.74 | 0.019 | 13874 | 62 | E |
| 7 12 10.97 | 55 22 30.4 | 15.55‡ | 0.250 | 11389 | 34 | N | 7 13 54.11 | 55 39 41.5 | 16.75 | 0.029 | 20668 | 44 | E |
| 7 20 51.66 | 55 13 45.0 | 15.57‡ | 0.250 | 11768 | 35 | N | 7 15 21.83 | 56 03 00.1 | 16.75 | 0.025 | 11937 | 48 | E |
| 7 19 25.65 | 55 39 47.9 | 15.58 | 0.030 | 11427 | 57 | E | 7 16 35.43 | 56 04 22.6 | 16.77 | 0.027 | 11351 | 52 | E |
| 7 18 23.22 | 55 25 40.8 | 15.59 | 0.034 | 12327 | 34 | N | 7 15 50.68 | 55 31 52.2 | 16.78 | 0.032 | 19620 | 43 | N |
| 7 20 17.66 | 55 07 20.6 | 15.60‡ | 0.250 | 11545 | 43 | N | 7 16 16.68 | 55 23 37.8 | 16.78 | 0.022 | 24884 | 68 | E |
| 7 13 08.97 | 55 57 26.9 | 15.61‡ | 0.250 | 30295 | 58 | N | 7 17 54.79 | 55 46 17.2 | 16.78 | 0.038 | 10252 | 50 | E |
| 7 15 57.31 | 55 37 21.8 | 15.61 | 0.031 | 12255 | 41 | N | 7 19 08.88 | 55 58 32.9 | 16.78 | 0.040 | 11296 | 39 | N |
| 7 18 00.53 | 56 11 35.2 | 15.61 | 0.017 | 10434 | 53 | E | 7 14 55.49 | 55 43 17.4 | 16.79 | 0.020 | 24800 | 70 | E |

## Velocities and Photometry continued

| RA (1950) | Decl | $R_{23.5}$ | $\sigma_R$ | $v$ | $\sigma_v$ | T | RA (1950) | Decl | $R_{23.5}$ | $\sigma_R$ | $v$ | $\sigma_v$ | T |
|---|---|---|---|---|---|---|---|---|---|---|---|---|---|
| 7 15 40.86 | 56 27 52.5 | 15.63‡ | 0.250 | 14936 | 42 | N | 7 18 36.68 | 55 38 51.5 | 16.79 | 0.031 | 10933 | 53 | N |
| 7 21 39.59 | 55 41 30.2 | 15.63‡ | 0.250 | 10444† | 100 | N | 7 15 05.89 | 56 19 13.3 | 16.81 | 0.041 | 23477 | 39 | N |
| 7 15 30.93 | 55 56 12.3 | 15.69 | 0.032 | 12593 | 34 | N | 7 18 35.76 | 55 47 52.5 | 16.83 | 0.024 | 11477 | 59 | N |
| 7 17 09.06 | 55 39 51.0 | 15.69 | 0.020 | 11831 | 31 | N | 7 18 58.67 | 55 55 48.6 | 16.83 | 0.040 | 11143 | 47 | N |
| 7 16 20.65 | 56 13 48.3 | 15.70 | 0.016 | 11034 | 37 | N | 7 15 41.18 | 55 39 18.2 | 16.86 | 0.033 | 9771 | 70 | E |
| 7 19 24.98 | 56 04 35.1 | 15.70 | 0.023 | 17184 | 41 | N | 7 16 02.77 | 55 32 12.7 | 16.87 | 0.032 | 20291 | 48 | N |
| 7 19 35.13 | 56 03 18.1 | 15.72 | 0.050 | 17592 | 40 | E | 7 17 30.13 | 55 56 52.8 | 16.87 | 0.022 | 10445 | 45 | N |
| 7 18 49.84 | 55 50 35.2 | 15.73 | 0.023 | 11932† | 100 | N | 7 15 10.55 | 55 29 01.0 | 16.88 | 0.044 | 18440 | 42 | E |
| 7 19 27.66 | 56 21 39.0 | 15.73 | 0.028 | 13118 | 49 | N | 7 16 22.10 | 56 20 25.9 | 16.88 | 0.018 | 11116 | 48 | E |
| 7 16 25.25 | 56 29 52.6 | 15.74‡ | 0.250 | 13429 | 47 | N | 7 17 31.61 | 56 20 49.7 | 16.88 | 0.018 | 57409 | 47 | N |
| 7 17 18.50 | 55 50 47.3 | 15.75 | 0.030 | 11483 | 36 | N | 7 14 23.09 | 55 25 24.8 | 16.91 | 0.038 | 23036 | 38 | E |
| 7 17 40.92 | 55 48 22.8 | 15.76 | 0.037 | 11382† | 100 | N | 7 20 38.26 | 55 23 59.4 | 16.92 | 0.038 | 70423 | 37 | N |
| 7 18 36.09 | 56 12 27.9 | 15.76 | 0.028 | 10225 | 45 | N | 7 18 32.39 | 55 40 28.0 | 16.93 | 0.038 | 12732 | 37 | N |
| 7 17 39.00 | 55 30 29.6 | 15.77 | 0.022 | 10805 | 35 | N | 7 16 11.35 | 55 26 19.8 | 16.95 | 0.023 | 19988 | 53 | N |
| 7 16 07.01 | 55 29 08.8 | 15.78 | 0.021 | 11431 | 33 | N | 7 16 36.41 | 55 25 00.9 | 16.95 | 0.021 | 25232 | 32 | N |
| 7 16 21.24 | 55 36 52.0 | 15.78 | 0.020 | 11000 | 51 | E | 7 19 46.12 | 55 50 07.1 | 16.95 | 0.024 | 11996 | 52 | E |
| 7 17 39.42 | 55 47 31.7 | 15.80 | 0.037 | 10688 | 40 | N | 7 16 04.89 | 56 02 50.0 | 16.99 | 0.027 | 10487 | 55 | N |
| 7 15 40.33 | 55 46 23.7 | 15.82 | 0.019 | 11545 | 46 | N | 7 19 37.92 | 56 18 15.9 | 16.99 | 0.029 | 12227 | 81 | N |
| 7 16 58.17 | 55 54 58.1 | 15.82 | 0.030 | 11561† | 100 | N | 7 15 45.82 | 56 03 26.0 | 17.00 | 0.025 | 71117 | 58 | N |
| 7 22 38.48 | 55 51 58.2 | 15.84‡ | 0.250 | 11279 | 52 | N | 7 13 38.13 | 55 21 21.5 | 17.07 | 0.037 | 50701 | 53 | N |
| 7 15 08.95 | 56 06 59.9 | 15.85 | 0.024 | 11336 | 36 | E | 7 18 38.61 | 56 09 19.2 | 17.15 | 0.023 | 11666 | 44 | N |
| 7 13 27.30 | 55 34 05.3 | 15.86‡ | 0.250 | 10269 | 44 | N | 7 15 01.00 | 55 24 12.2 | 17.16 | 0.044 | 19259 | 69 | E |
| 7 17 52.76 | 55 55 00.2 | 15.86 | 0.021 | 12680† | 100 | N | 7 15 38.50 | 55 20 33.7 | 17.17 | 0.043 | 31922 | 37 | E |
| 7 15 02.10 | 55 35 06.5 | 15.87 | 0.031 | 19633 | 46 | N | 7 15 30.38 | 55 47 35.6 | 17.21 | 0.021 | 24961 | 27 | N |
| 7 19 39.15 | 56 10 04.5 | 15.88 | 0.023 | 12052 | 50 | N | 7 17 33.33 | 56 00 03.0 | 17.22 | 0.022 | 10577 | 55 | N |
| 7 18 50.38 | 55 43 55.9 | 15.89 | 0.023 | 14441† | 100 | N | 7 15 41.89 | 55 22 14.0 | 17.23 | 0.044 | 31475 | 68 | E |
| 7 17 24.91 | 56 20 48.8 | 15.90 | 0.017 | 10970 | 41 | N | 7 14 05.78 | 56 00 09.4 | 17.27 | 0.058 | 10937 | 80 | E |
| 7 19 41.45 | 55 48 28.7 | 15.90 | 0.023 | 11384† | 100 | N | 7 15 49.59 | 55 38 42.7 | 17.37 | 0.032 | 11814 | 47 | N |
| 7 18 43.53 | 55 59 22.2 | 15.91 | 0.040 | 11083† | 100 | N | 7 17 22.02 | 55 26 39.3 | 17.42 | 0.036 | 19758† | 100 | N |
| 7 21 18.89 | 56 05 34.0 | 15.91‡ | 0.250 | 19658 | 46 | N | 7 17 47.60 | 55 48 23.1 | 17.45 | 0.039 | 30601 | 41 | N |
| 7 19 18.55 | 55 25 51.1 | 15.92 | 0.033 | 12480 | 42 | N | 7 17 02.76 | 55 50 42.8 | 17.63 | 0.034 | 11653 | 71 | E |
| 7 18 34.41 | 56 26 14.3 | 15.93‡ | 0.250 | 16937 | 66 | E | 7 17 14.67 | 55 55 00.4 | 17.67 | 0.033 | 13002 | 63 | N |
| 7 14 39.43 | 55 46 21.5 | 15.94 | 0.028 | 11688 | 55 | E | 7 16 59.25 | 56 18 56.5 | 17.70 | 0.021 | 23205 | 68 | E |
| 7 15 52.01 | 55 16 05.2 | 15.94‡ | 0.250 | 31444 | 39 | E | 7 19 53.25 | 56 11 30.7 | 17.73 | 0.039 | 23847 | 71 | E |
| 7 14 50.17 | 56 14 38.0 | 15.96 | 0.041 | 10995 | 43 | E | 7 15 14.58 | 55 30 43.2 | 17.74 | 0.034 | 19372 | 54 | N |
| 7 14 52.41 | 55 26 45.5 | 15.96 | 0.043 | 14259 | 52 | E | 7 20 59.79 | 55 27 54.6 | 17.74 | 0.041 | 9475 | 69 | E |
| 7 17 13.42 | 55 53 02.3 | 15.96 | 0.030 | 11787† | 100 | N | 7 14 23.46 | 55 27 56.7 | 17.76 | 0.038 | 17910 | 70 | E |
| 7 22 16.05 | 56 25 46.5 | 15.96‡ | 0.250 | 11860 | 46 | N | 7 16 54.05 | 55 25 37.1 | 17.80 | 0.028 | 60396 | 58 | N |
| 7 16 00.38 | 55 35 15.9 | 15.97 | 0.031 | 11935 | 75 | E | 7 19 21.76 | 55 21 15.9 | 17.84 | 0.036 | 36839 | 60 | N |
| 7 13 07.99 | 55 15 11.9 | 16.03‡ | 0.250 | 11590 | 62 | E | 7 16 35.87 | 56 05 27.8 | 17.86 | 0.029 | 30303 | 47 | E |
| 7 17 05.38 | 56 29 47.7 | 16.03‡ | 0.250 | 10961 | 48 | E | 7 20 02.99 | 55 51 10.6 | 17.86 | 0.042 | 17301 | 57 | E |
| 7 19 16.41 | 55 34 57.5 | 16.03 | 0.030 | 5606 | 38 | N | 7 17 26.07 | 55 50 10.9 | 18.06 | 0.042 | 11573 | 51 | N |
| 7 14 23.33 | 55 35 59.3 | 16.04 | 0.039 | 11813 | 51 | E | | | | | | | |

† Hintzen *et al.* 1982  
× Hill *et al.* 1980  
‡ POSS digitized scan photometry  
\* Marzke & Huchra 1995